\documentclass{ws-ijmpa}
\usepackage[super,compress]{cite}
\usepackage{graphicx}

\usepackage{epsfig}
\usepackage{epstopdf}
\usepackage{color}

\begin{document}
\markboth{Taras V. Zagoskin and Alexander Yu. Korchin}{Higgs decay into $ZZ$ decaying to identical fermions}

%
\catchline{}{}{}{}{}
%

\title{The Higgs boson decay into $ZZ$ decaying to identical fermion pairs}

\author{Taras~V.~Zagoskin}

\address{Institute of Theoretical Physics, NSC ``Kharkov Institute of Physics and Technology'', \\
Kharkov, 61108 Ukraine\\
taras.zagoskin@gmail.com}

\author{Alexander~Yu.~Korchin}

\address{Institute of Theoretical Physics, NSC ``Kharkov Institute of Physics and Technology'', \\
Kharkov, 61108 Ukraine\\
korchin@kipt.kharkov.ua}

\maketitle

\begin{history}
\received{Day Month Year}
\revised{Day Month Year}
\end{history}

\begin{abstract}
In order to investigate various decay channels of the Higgs boson $h$ or the hypothetical dilaton, we consider a neutral particle $X$ with zero spin and arbitrary $CP$ parity. This particle can decay into two off-mass-shell $Z$ bosons ($Z_1^*$ and $Z_2^*$) decaying to identical fermion-antifermion pairs ($f \bar{f}$): $X \to Z_1^* Z_2^* \to f \bar{f} f \bar{f}$. We derive analytical formulas for the fully differential width of this decay and for the fully differential width of $h \to Z_1^* Z_2^* \to 4 \ell$ ($4 \ell$ stands for $4 e$, $4 \mu$, or $2 e 2 \mu$). Integration of these formulas yields some Standard Model histogram distributions of the decay $h \to Z_1^* Z_2^* \to 4 \ell$ which are compared with corresponding Monte Carlo simulated distributions obtained by ATLAS and with ATLAS experimental data.

\end{abstract}

\ccode{PACS numbers: 12.15.Ji, 12.60.Fr, 14.80.Bn, 14.80.Ec}

\section{Introduction}
\label{Section: Introduction}

The boson $h$ discovered \cite{Aad:2012, Chatrchyan:2012} in 2012 by the CMS and ATLAS collaborations was reported to have a mass about 125 GeV and some decay modes predicted for the Standard Model (SM) Higgs boson. Since that time, the observed particle, called the Higgs boson, has been intensively studied (see, for example, Refs.~\citen{Aad, Hue, Merchand, Kumar, Kawasaki, Pallis, Baglio, Evnin, Maroto, Groeber, Romero, Chen, Basirnia, Barbieri, Angelescu, Prilepina, Dorsch, Haarr, Spannowsky, Carpenter, Englert, Berge, Gritsan:2016hjl, He:2014xla, Hajer:2015iul}). A main goal of experiments on the Higgs boson physics has been to prove or disprove the hypothesis that $h$ is the SM Higgs boson. Apart from the decay channels, the SM predicts that $h$ has $J^{CP} = 0^{++}$. The followed thorough analysis has fine-tuned the mass of $h$, which is $125.09 \pm 0.24$ GeV according to Ref.~\citen{PDG:2015}, and has yielded some information on its spin and its $CP$ parity.

In particular, the observation of the $h \to ZZ$ and $h \to W^- W^+$ modes (see, for example, Ref.~\citen{Aad:2016}) means that the Higgs boson spin is zero, one, or two while the fact that $h$ decays \cite{Aad:2016} to $\gamma \gamma$ and the Landau-Yang theorem exclude the spin-one variant. Further, the analyses presented in Ref.~\citen{Aad:2015, Khachatryan:2015} rule out many spin-two hypotheses at a 99\% confidence level (CL) or higher. Therefore, we conclude that the spin of the Higgs boson is zero with a probability of about 99\%.

To clarify the $CP$ properties of $h$, in Ref.~\citen{Zagoskin:2016} we study the decay of a spin-zero particle $X$ into two off-mass-shell $Z$ bosons $Z_1^*$ and $Z_2^*$. Since $X$ is defined as an elementary neutral particle with zero spin, our study applies to the Higgs boson. Moreover, it can apply to the dilaton if this boson actually exists.

The amplitude of the decay $X \to Z_1^* Z_2^*$ depends (see Eq.~(4) in Ref.~\citen{Zagoskin:2016}) on 3 complex-valued functions of the invariant masses of $Z_1^*$ and $Z_2^*$. These functions determine the $CP$ properties of the boson $X$ and are called the $XZZ$ couplings. Using the CMS and ATLAS experimental data on the decay $h \to Z_1^* Z_2^* \to 4 \ell$ (where $4 \ell$ stands for $4 e$, $4 \mu$, or $2 e 2 \mu$), these collaborations in Refs.~\citen{Aad:2015, Khachatryan:2015, Aad:2016} and we in Ref.~\citen{Zagoskin:2016} have obtained some constraints on the $hZZ$ couplings. These constraints demonstrate that $h$ is not a $CP$-odd state and it may be the SM Higgs boson, another $CP$-even state, or a boson with indefinite $CP$ parity. Besides, as shown in Ref.~\citen{Zagoskin:2016}, a non-zero imaginary part of the $hZZ$ couplings is not excluded, which can be related to small loop corrections and possibly to a non-Hermiticity of the $hZZ$ interaction.

Thus, the $CP$ parity of the Higgs boson is not yet fully ascertained. Moreover, in some supersymmetric extensions of the SM there are \cite{Pilaftsis:1999np,Barger:2009pr,Branco:2012pre} neutral bosons with negative or indefinite $CP$ parity. That is why it is now important to establish the $CP$ properties of the Higgs boson.

Aiming at that, we consider the decay of the particle $X$ into $Z_1^*$ and $Z_2^*$ which then decay to fermion-antifermion pairs $f_1 \bar{f}_1$ and $f_2 \bar{f}_2$ respectively. While in Ref.~\citen{Zagoskin:2016} we study in detail the decays with the non-identical fermions, $f_1 \neq f_2$, in the present paper the case $f_1 = f_2$ is under investigation. The masses of the fermions $f_1$ and $f_2$ are neglected in both papers.

We are motivated to consider the decay into identical fermions by the following. In Refs.~\citen{Khachatryan:2015, Aad:2015} the CMS and ATLAS collaborations analyze 95 events $h \to Z_1^* Z_2^* \to 4 \ell$. 53 of them are the decays to identical leptons, namely to $4 e$ or $4 \mu$. In spite of the fact that the decays to the identical leptons make up about 55\% of the measured decays $h \to Z_1^* Z_2^* \to 4 \ell$, the distributions of the former decays have not been properly analytically studied.

The SM total widths of the decays into identical fermions are studied in Refs.~\citen{Bredenstein:2006, Dittmaier:2011} and are calculated in Ref.~\citen{LHC_data:twiki}. Some distributions of the decay $X \to Z_1^* Z_2^* \to 4 \ell$ are plotted in Ref.~\citen{Aad:2015, Khachatryan:2015} for the SM Higgs boson and some spin-zero states beyond the SM.
In the present paper we perform a more general study and consider the decay $X \rightarrow Z_1^* Z_2^* \rightarrow f \bar{f} f \bar{f}$ with allowance for all the possible $CP$ properties of the particle $X$.

In Sec.~\ref{Section: The fully differential width} we derive an analytical formula for the fully differential width of the decay to identical fermions. Section~\ref{Section: Invariant mass and angular distributions} shows a comparison of some distributions of the decay to identical leptons with those for the decay into non-identical ones. For this comparison we obtain an exact analytical formula for a certain differential width of the decay to non-identical fermions (see \ref{Appendix: A formula for d Gamma over d a of the decay into non-identical fermions}). We analyze the usefulness of all the compared distributions for obtaining constraints on the $hZZ$ couplings. In Sec.~\ref{Section: Comparison with experimental data} we derive some SM histogram distributions of the decay $h \to Z_1^* Z_2^* \to 4 \ell$ by Monte Carlo (MC) integration and compare them with the corresponding simulations presented in Ref.~\citen{Aad:2015} and with the experimental distributions from Ref.~\citen{Aad:2015}.

\section{The fully differential width}
\label{Section: The fully differential width}

We consider a neutral particle $X$ with zero spin and arbitrary $CP$ parity. It can decay into two fermion-antifermion pairs, $f_1 \bar{f}_1$ and $f_2 \bar{f}_2$, through the two off-mass-shell $Z$ bosons ($Z_1^*$ and $Z_2^*$):
\begin{align}
\label{X-> Z_1^* Z_2^* -> f_1 antif_1 f_2 antif_2}
X \rightarrow Z_1^* Z_2^* \rightarrow f_1 \bar{f}_1 f_2 \bar{f}_2.
\end{align}
If $m_X \in (4 m_b, 2 m_t]$ ($m_X$ is the mass of the particle $X$, $m_b$ is the mass of the $b$ quark, $m_t$ is the mass of the $t$ quark), which holds for $X = h$, then $f_j = e^-, \mu^-, \tau^-, \nu_e, \nu_{\mu}, \nu_{\tau}, u, c, d, s, b$. If $m_X > 4 m_t$, which is possible \cite{Gasperini:1994} if $X$ is the dilaton, then $f_j$ can be the top quark as well.

In Ref.~\citen{Zagoskin:2016} we considered decays
\begin{align}
\label{X-> Z_1^* Z_2^* -> f_1 antif_1 f_2 antif_2, f_1 neq f_2}
X \rightarrow Z_1^* Z_2^* \rightarrow f_1 \bar{f}_1 f_2 \bar{f}_2, \qquad f_1 \neq f_2
\end{align}
at the tree level.

\begin{figure}[h]
\begin{minipage}[h]{0.47 \linewidth}
\center{\includegraphics[scale=0.55]{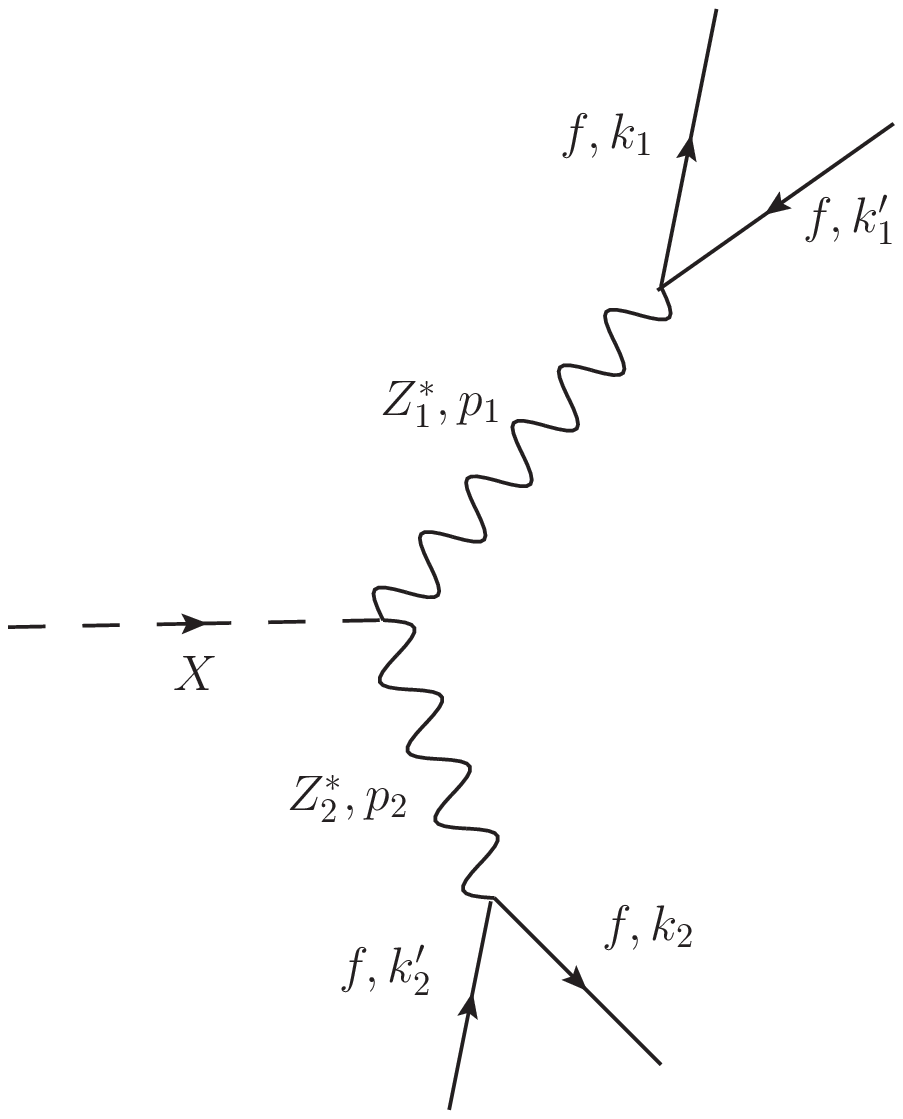} \\ a}
\end{minipage}
\hfill
\begin{minipage}[h]{0.47 \linewidth}
\center{\includegraphics[scale=0.55]{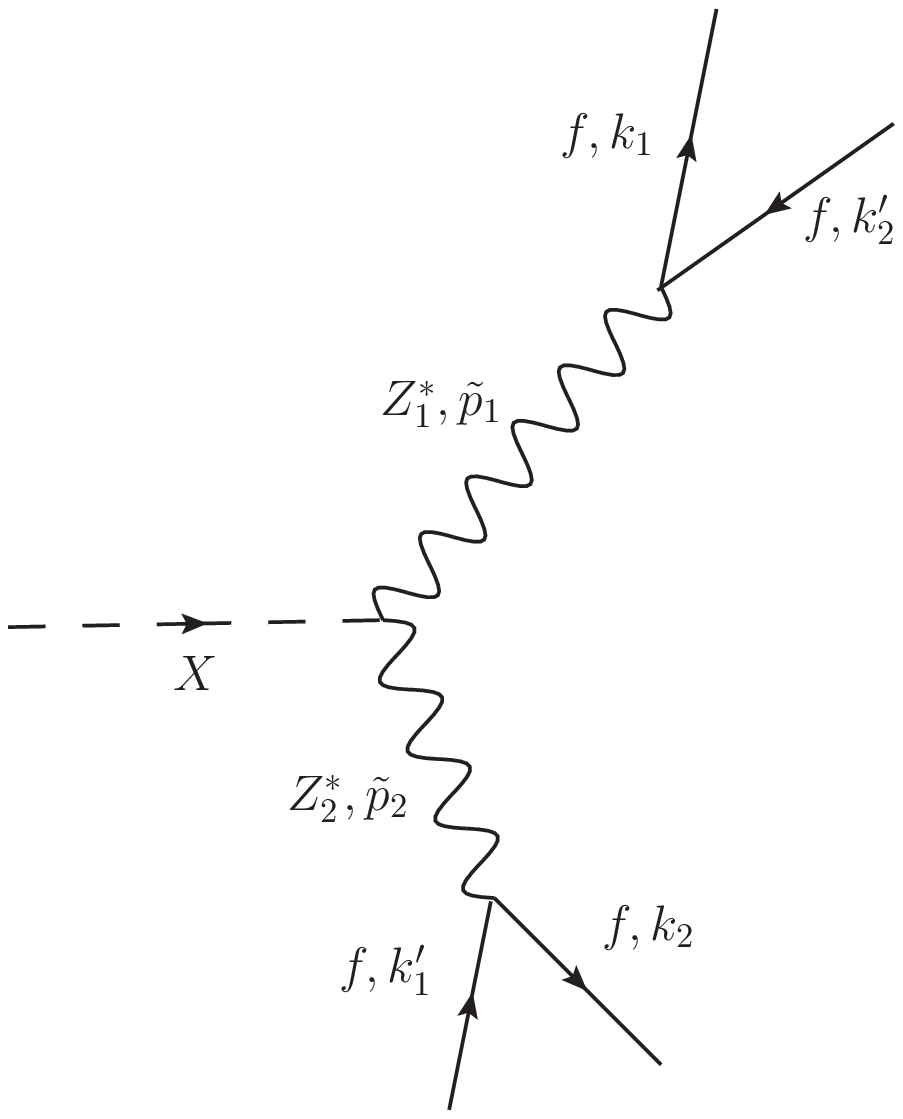} \\ b}
\end{minipage}
\caption{The Feynman diagrams contributing to the matrix element of decay (\ref{X-> Z_1^* Z_2^* -> f antif f antif}).}
\label{The Feynman diagrams for the decay into identical fermions}
\end{figure}

The present paper shows our analysis of decay (\ref{X-> Z_1^* Z_2^* -> f_1 antif_1 f_2 antif_2}) in the case of the identical fermions, $f_1 = f_2 \equiv f$:
\begin{align}
\label{X-> Z_1^* Z_2^* -> f antif f antif}
X \rightarrow Z_1^* Z_2^* \rightarrow f \bar{f} f \bar{f}.
\end{align}
The matrix element of decay (\ref{X-> Z_1^* Z_2^* -> f antif f antif}) is
\begin{align}
\label{The matrix element of the decay to identical fermions}
M_{iden} = M - \tilde{M},
\end{align}
where the matrix elements $M$ and $\tilde{M}$ correspond to the diagrams (a) and (b) in Fig.~\ref{The Feynman diagrams for the decay into identical fermions} respectively. Namely,

%

%
\begin{align}
\label{Expressions for M and tilde M}
M = & \frac{i}{(a_1 - m_Z^2 + i m_Z \Gamma_Z) (a_2 - m_Z^2 + i m_Z \Gamma_Z)} \sum_{\lambda_1, \lambda_2 = -1, 0, 1} A_{X \to Z_1^* Z_2^*} (p_1, p_2, \lambda_1, \lambda_2) & \notag \\
& \times A_{Z \to f \bar{f}} (k_1, k_1^{\prime}, \lambda_{f_1}, \lambda_{\bar{f}_1}, \lambda_1) A_{Z \to f \bar{f}} (k_2, k_2^{\prime}, \lambda_{f_2}, \lambda_{\bar{f}_2}, \lambda_2), & \notag \\
\tilde{M} = & \frac{i}{(\tilde{a}_1 - m_Z^2 + i m_Z \Gamma_Z) (\tilde{a}_2 - m_Z^2 + i m_Z \Gamma_Z)} \sum_{\lambda_1, \lambda_2 = -1, 0, 1} A_{X \to Z_1^* Z_2^*} (\tilde{p}_1, \tilde{p}_2, \lambda_1, \lambda_2) & \notag \\
& \times A_{Z \to f \bar{f}} (k_1, k_2^{\prime}, \lambda_{f_1}, \lambda_{\bar{f}_2}, \lambda_1) A_{Z \to f \bar{f}} (k_2, k_1^{\prime}, \lambda_{f_2}, \lambda_{\bar{f}_1}, \lambda_2), &
\end{align}
where
\begin{itemize}
\item $k_1$ and $k_1^{\prime}$ ($k_2$ and $k_2^{\prime}$) are the 4-momenta of the particles $f_1$ and $\bar{f}_1$ ($f_2$ and $\bar{f}_2$) in the rest frame of $X$;
\item $p_1 \equiv k_1 + k_1^{\prime}$ and $p_2 \equiv k_2 + k_2^{\prime}$ are the 4-momenta of $Z_1^*$ and $Z_2^*$ respectively in the rest frame of $X$ in diagram Fig.~\ref{The Feynman diagrams for the decay into identical fermions}~(a);
\item $a_j \equiv p_j^2$;
\item $m_Z$ and $\Gamma_Z$ are respectively the pole mass and the total width of the $Z$ boson;
\item $A_{X \to Z_1^* Z_2^*} (p_1, p_2, \lambda_1, \lambda_2)$ is the amplitude of the decay $X \to Z_1^* Z_2^*$ where $p_j$ and $\lambda_j$ are respectively the momentum and the helicity of the boson $Z_j^*$ in the rest frame of $X$;
\item $A_{Z \to f \bar{f}} (k, k^{\prime}, \lambda_f, \lambda_{\bar{f}}, \lambda)$ is the amplitude of the decay $Z \to f \bar{f}$ where $k$ and $\lambda_f$ ($k^{\prime}$ and $\lambda_{\bar{f}}$) are respectively the momentum and the polarization of $f$ ($\bar{f}$) in the rest frame of $Z$, $\lambda$ is the helicity of decaying $Z$;
\item $\tilde{p}_1 \equiv k_1 + k_2^{\prime}$ and $\tilde{p}_2 \equiv k_2 + k_1^{\prime}$ are the 4-momenta of $Z_1^*$ and $Z_2^*$ respectively in the rest frame of $X$ in diagram Fig.~\ref{The Feynman diagrams for the decay into identical fermions}~(b);
\item $\tilde{a}_j \equiv \tilde{p}_j^2$.
\end{itemize}

From the conservation of the energy-momentum 4-vectors we find all the possible values of $a_1$ and $a_2$:
\begin{equation}
\label{intervals of a_1 and a_2}
4 m_{f_1}^2 < a_1 <  (m_X - 2 m_{f_2})^2, \qquad 4 m_{f_2}^2 < a_2 < (m_X - \sqrt{a_1})^2,
\end{equation}
where $m_{f_j}$ is the mass of the fermion $f_j$.

The amplitude $A_{X \to Z_1^* Z_2^*} (p_1, p_2, \lambda_1, \lambda_2)$ is  \cite{Zagoskin:2016}
\begin{align}
\label{the amplitude of X->Z_1* Z_2* for the unpermuted diagram}
A_{X \rightarrow Z_1^* Z_2^*} (p_1, p_2, \lambda_1, \lambda_2) = & g_Z \Biggl (a_Z (a_1, a_2)  (e_1^* \cdot e_2^*) + \frac {b_Z (a_1, a_2)} {m_X^2} (e_1^* \cdot p_X) (e_2^* \cdot p_X) & \notag \\
& + i \frac {c_Z (a_1, a_2)} {m_X^2} \varepsilon_{\mu \nu \rho \sigma} p_X^{\mu} (p_1^{\nu} - p_2^{\nu}) (e_1^{\rho})^* (e_2^{\sigma})^* \Biggr), &
\end{align}
where $g_Z \equiv 2 \sqrt {\sqrt{2} G_F} m_Z^2$, $G_F$ is the Fermi constant, $a_Z (a_1, a_2)$, $b_Z (a_1, a_2)$, and $c_Z (a_1, a_2)$ are some complex-valued dimensionless functions of $a_1$ and $a_2$, $e_j \equiv e (p_j, \lambda_j)$ with $e (p, \lambda)$ being the polarization 4-vector of the $Z$ boson with a momentum $p$ and a helicity $\lambda$, $p_X \equiv p_1 + p_2 = \tilde{p}_1 + \tilde{p}_2 = (m_X, \vec{0})$ is the 4-momentum of the boson $X$ in its own rest frame, $\varepsilon_{\mu \nu \rho \sigma}$ is the Levi-Civita symbol ($\varepsilon_{0123} = 1$).

The values of the couplings $a_Z$, $b_Z$, and $c_Z$ reflect the $CP$ properties of the particle $X$. Specifically, at the tree level the correspondence shown in Table~\ref{The CP parity of X in the case of various sets of values of a_Z, b_Z, and c_Z} takes place.

\begin{table}[h]
\tbl{The $CP$ parity of the particle $X$ for various values of $a_Z$, $b_Z$, and $c_Z$.}
{\label{The CP parity of X in the case of various sets of values of a_Z, b_Z, and c_Z}
\begin{tabular}[t]{c | c  c  c}
\hline
$CP_X$ & $a_Z$ & $b_Z$ & $c_Z$ \\
\hline
1 & {\rm any} & {\rm any} & 0 \\
\hline
$-1$ & 0 & 0 & $\neq$ 0 \\
\hline
{\rm indefinite} & $\neq$ 0 & {\rm any} & $\neq$ 0\\
& {\rm any} & $\neq 0$ & $\neq 0$ \\
\hline
\end{tabular}}
\end{table}

For the SM Higgs boson the loop corrections change slightly the tree-level values $a_Z = 1$, $b_Z = 0$, $c_Z = 0$ (see, for example, Refs.~\citen{Bolognesi:2012, Anderson:2013, Khachatryan:2015, Soni:1993}). In particular, the SM electroweak radiative diagrams tune the value of the coupling $b_Z$, beginning from the next-to-leading order, while a contribution to $c_Z$ appears at the three-loop level, so that $|b_Z| \approx 10^{-2}$ and $|c_Z| \approx 10^{-11}$ (see Ref.~\citen{Gao:2010qx}). Physics beyond the SM is the additional source of a possible deviation from the values $a_Z = 1$, $b_Z = 0$, $c_Z = 0$.

Calculating Lorentz-invariant amplitude (\ref{the amplitude of X->Z_1* Z_2* for the unpermuted diagram}) in the rest frame of $X$, we derive that
\begin{align}
\label{The amplitudes of X->Z_1* Z_2* for the unpermuted diagram at various helicities of Z_1* and Z_2*}
& A_{X \rightarrow Z_1^* Z_2^*} (p_1, p_2, \pm 1, \pm 1) = g_Z \left (a_Z (a_1, a_2) \pm c_Z (a_1, a_2) \frac{k}{m_X^2} \right), & \notag \\
& A_{X \rightarrow Z_1^* Z_2^*} (p_1, p_2, 0, 0) = - g_Z \left (a_Z (a_1, a_2) \frac{m_X^2 - a_1 - a_2}{2 \sqrt{a_1 a_2}} + b_Z (a_1, a_2) \frac{k^2}{4 m_X^2 \sqrt{a_1 a_2}} \right), & \notag \\
& A_{X \rightarrow Z_1^* Z_2^*} (p_1, p_2, \lambda_1, \lambda_2) = 0, \qquad \lambda_1 \neq \lambda_2, &
\end{align}
where $k (a_1, a_2) \equiv \lambda^{1/2} (m_X^2, a_1, a_2)$, $\lambda (x, y, z) \equiv x^2 + y^2 + z^2 - 2xy - 2xz - 2 yz$.

We take the amplitude $A_{Z \to f \bar{f}} (k, k^{\prime}, \lambda_f, \lambda_{\bar{f}}, \lambda)$ from the SM (see, for example, Ref.~\citen{Novikov:1999}).

\begin{figure}[h]
\includegraphics[scale=0.7]{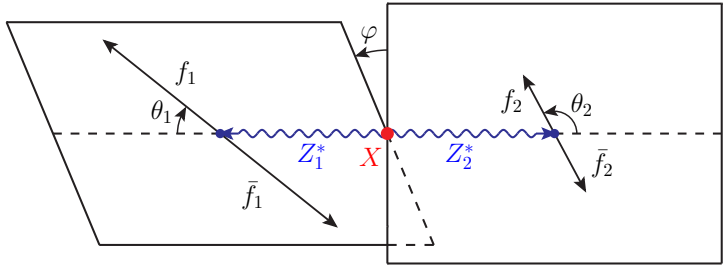}
\caption{The kinematics of decay (\ref{X-> Z_1^* Z_2^* -> f_1 antif_1 f_2 antif_2}). We show the momenta of $Z_1^*$ and $Z_2^*$ in the rest frame of $X$ while the momenta of $f_1$ and $\bar{f}_1$ ($f_2$ and $\bar{f}_2$) are shown in the rest frame of $Z_1^*$ ($Z_2^*$).}
\label{The momenta of the particles produced in a decay X to ZZ to 4f}
\end{figure}

Further, to describe decay (\ref{X-> Z_1^* Z_2^* -> f_1 antif_1 f_2 antif_2}), let us introduce the following angles (see Fig.~\ref{The momenta of the particles produced in a decay X to ZZ to 4f}): $\theta_1$ ($\theta_2$) is the angle between the momentum of $Z_1^*$ ($Z_2^*$) in the rest frame of $X$ and the momentum of $f_1$ ($f_2$) in the rest frame of $Z_1^*$ ($Z_2^*$) (in other words, $\theta_1$ ($\theta_2$) is the polar angle of the fermion $f_1$ ($f_2$)) and $\varphi$ is the azimuthal angle between the planes of the decays $Z_1^* \rightarrow f_1 \bar{f}_1$ and $Z_2^* \rightarrow f_2 \bar{f}_2$. For decay (\ref{X-> Z_1^* Z_2^* -> f antif f antif}), we can arbitrarily choose the $Z$ boson which we will call $Z_1^*$, and then we will refer to the other $Z$ boson as $Z_2^*$.

As for $\tilde{a}_1$ and $\tilde{a}_2$, an explicit calculation yields
\begin{align}
\label{Expressions for tilde a_1 and tilde a_2}
\tilde{a}_1 = \frac{m_X^2 - a_1 - a_2}{4} (1 - \cos \theta_1 \cos \theta_2) + \frac{\sqrt{a_1 a_2}}{2} \sin \theta_1 \sin \theta_2 \cos \phi + \frac{k}{4} (\cos \theta_1 - \cos \theta_2), & \notag \\
\tilde{a}_2 = \frac{m_X^2 - a_1 - a_2}{4} (1 - \cos \theta_1 \cos \theta_2) + \frac{\sqrt{a_1 a_2}}{2} \sin \theta_1 \sin \theta_2 \cos \phi + \frac{k}{4} (\cos \theta_2 - \cos \theta_1). &
\end{align}

The expression for the amplitude $A_{X \to Z_1^* Z_2^*} (\tilde{p}_1, \tilde{p}_2, \lambda_1, \lambda_2)$ is analogous to Eq.~(\ref{the amplitude of X->Z_1* Z_2* for the unpermuted diagram}):
\begin{align}
\label{the amplitude of X->Z_1* Z_2* for the permuted diagram}
A_{X \rightarrow Z_1^* Z_2^*} (\tilde{p}_1, \tilde{p}_2, \lambda_1, \lambda_2) = & g_Z \Biggl (a_Z (\tilde{a}_1, \tilde{a}_2)  (\tilde{e}_1^* \cdot \tilde{e}_2^*) + \frac {b_Z (\tilde{a}_1, \tilde{a}_2)} {m_X^2} (\tilde{e}_1^* \cdot p_X) (\tilde{e}_2^* \cdot p_X) & \notag \\
& + i \frac {c_Z (\tilde{a}_1, \tilde{a}_2)} {m_X^2} \varepsilon_{\mu \nu \rho \sigma} p_X^{\mu} (\tilde{p}_1^{\nu} - \tilde{p}_2^{\nu}) (\tilde{e}_1^{\rho})^* (\tilde{e}_2^{\sigma})^* \Biggr), &
\end{align}
where $\tilde{e}_j = e (\tilde{p}_j, \lambda_j)$. Calculating $A_{X \to Z_1^* Z_2^*} (\tilde{p}_1, \tilde{p}_2, \lambda_1, \lambda_2)$ in the rest frame of $X$, we get
\begin{align}
\label{The amplitudes of X->Z_1* Z_2* for the permuted diagram at various helicities of Z_1* and Z_2*}
& A_{X \rightarrow Z_1^* Z_2^*} (\tilde{p}_1, \tilde{p}_2, \pm 1, \pm 1) = g_Z \left (a_Z (\tilde{a}_1, \tilde{a}_2) \pm c_Z (\tilde{a}_1, \tilde{a}_2) \frac{2}{m_X} |\mathbf{k}_1 + \mathbf{k}_2^{\prime}| \right), & \notag \\
& A_{X \rightarrow Z_1^* Z_2^*} (\tilde{p}_1, \tilde{p}_2, 0, 0) = - \frac{g_Z} {4 \sqrt{\tilde{a}_1 \tilde{a}_2}} \Biggl (a_Z (\tilde{a}_1, \tilde{a}_2) \Bigl (m_X^2 + a_1 + a_2 + (m_X^2 - a_1 - a_2) & \notag \\
& \times \cos \theta_1 \cos \theta_2 - 2 \sqrt{a_1 a_2} \sin \theta_1 \sin \theta_2 \cos \phi \Bigr) + b_Z (\tilde{a}_1, \tilde{a}_2) \cdot 4 |\mathbf{k}_1 + \mathbf{k}_2^{\prime}|^2 \Biggr), & \notag \\
& A_{X \rightarrow Z_1^* Z_2^*} (\tilde{p}_1, \tilde{p}_2, \lambda_1, \lambda_2) = 0, \qquad \lambda_1 \neq \lambda_2, &
\end{align}
where
\begin{align}
|\mathbf{k}_1 + \mathbf{k}_2^{\prime}|^2 = & \frac{a_1 + a_2}{4} - \frac{\sqrt{a_1 a_2}}{2} \sin \theta_1 \sin \theta_2 \cos \phi + \frac{k^2}{16 m_X^2} (\cos^2 \theta_1 + \cos^2 \theta_2) & \notag \\
& + \frac{\cos \theta_1 \cos \theta_2}{8 m_X^2} (m_X^4 - (a_1 - a_2)^2). &
\end{align}

Using Eqs.~(\ref{The matrix element of the decay to identical fermions}), (\ref{Expressions for M and tilde M}), (\ref{The amplitudes of X->Z_1* Z_2* for the unpermuted diagram at various helicities of Z_1* and Z_2*}), (\ref{Expressions for tilde a_1 and tilde a_2}), and (\ref{The amplitudes of X->Z_1* Z_2* for the permuted diagram at various helicities of Z_1* and Z_2*}), we derive Eq.~(\ref{The fully differential width}) (see \ref{Appendix: A formula for the fully differential width of the decay into identical fermions}).

\section{Invariant mass and angular distributions}
\label{Section: Invariant mass and angular distributions}

\begin{figure}[h]

\center{\includegraphics[scale=0.45]{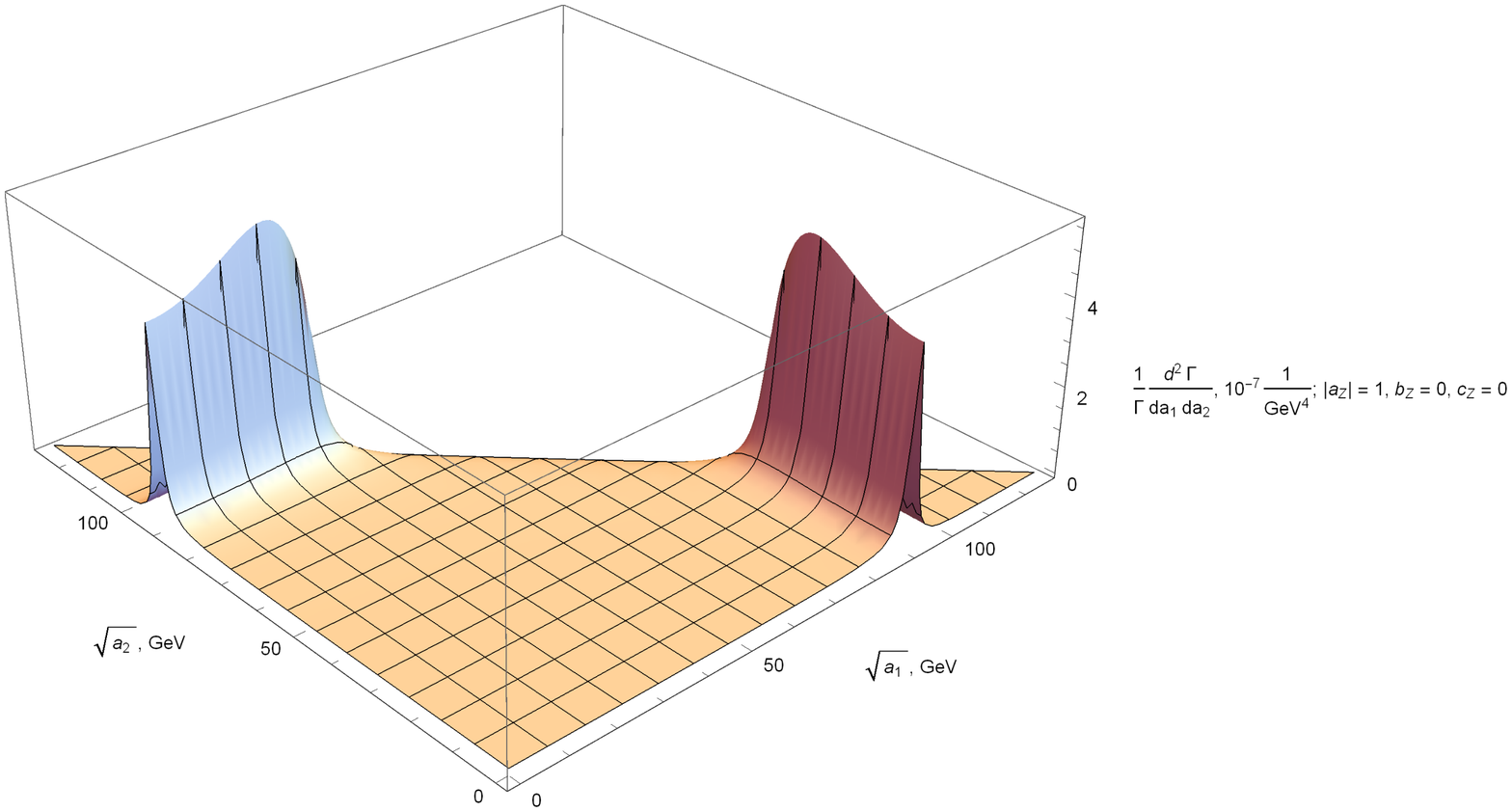} \\ a}
\center{\includegraphics[scale=0.45]{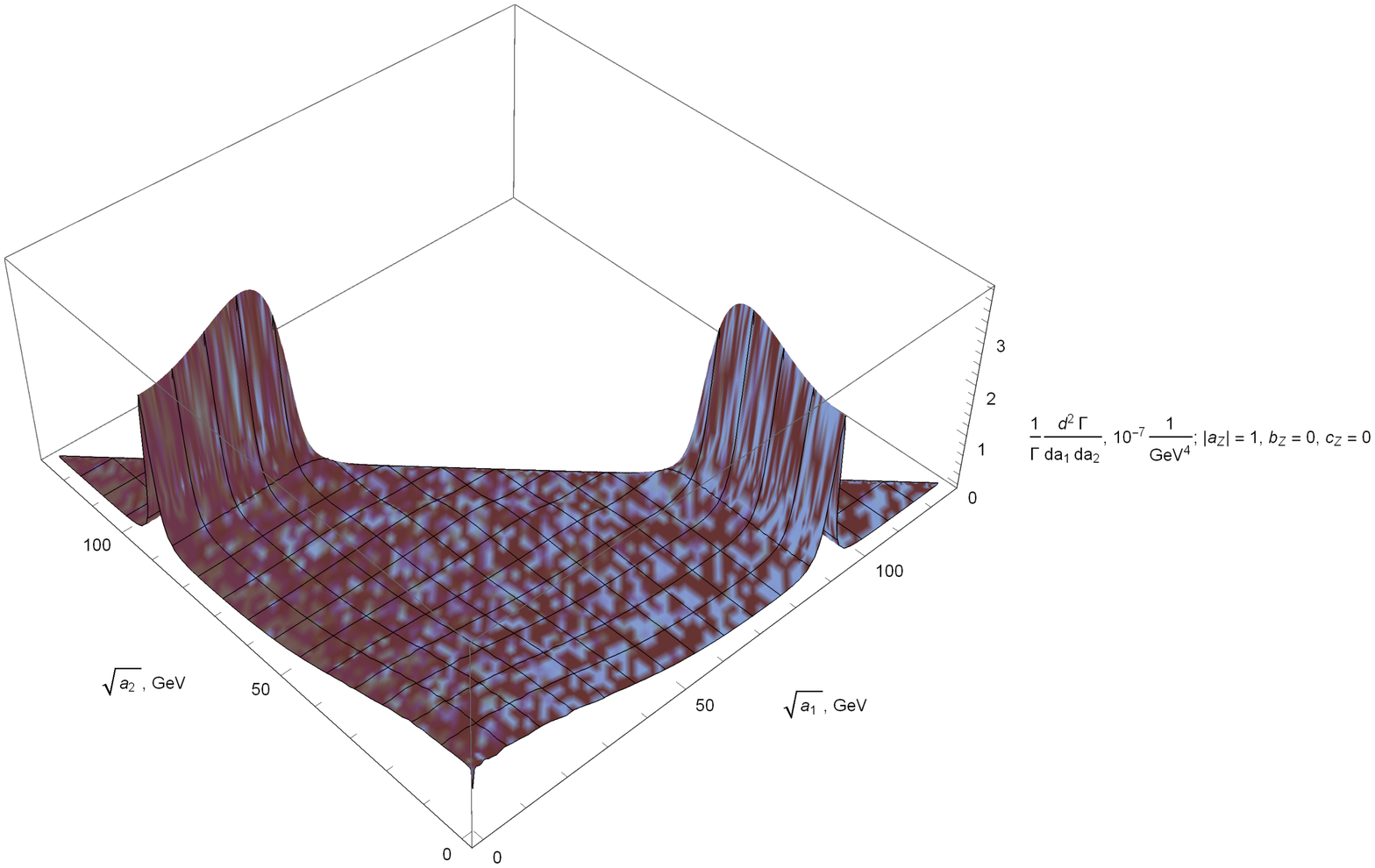} \\ b}

\caption{The distribution $\frac{1}{\Gamma} \frac{d^2 \Gamma}{d a_1 d a_2}$ (in units of $10^{-7} \, {\rm GeV}^{-4}$) in the SM for the decays $h \to Z_1^* Z_2^* \to f_1 \bar{f}_1 f_2 \bar{f}_2$ with $f_1 \neq f_2$ (a) and for the decays $h \to Z_1^* Z_2^* \to 4 l$ with $l = e, \mu, \tau$ (b).}
\label{Plots of the distributions of a1 and a2}
\end{figure}

Integrating Eq.~(\ref{The fully differential width}) numerically, we can obtain some distributions of decay (\ref{X-> Z_1^* Z_2^* -> f antif f antif}). Moreover, numerical integration of Eq.~(5) in Ref.~\citen{{Zagoskin:2016}} yields distributions for decay (\ref{X-> Z_1^* Z_2^* -> f_1 antif_1 f_2 antif_2, f_1 neq f_2}). In Figs.~\ref{Plots of the distributions of a1 and a2} and \ref{Plots of the distributions of a, theta, and phi} we compare certain distributions of (\ref{X-> Z_1^* Z_2^* -> f antif f antif}) with those of (\ref{X-> Z_1^* Z_2^* -> f_1 antif_1 f_2 antif_2, f_1 neq f_2}). We define the weak mixing angle as $\theta_W \equiv \arcsin \sqrt{1 - m_W^2 / m_Z^2}$, where $m_W$ is the mass of the $W$ boson, and use the values of the constants in Table~\ref{The values of G_F, m_h, m_Z, m_W, and Gamma_Z according to the 2014 data of Particle Data Group} neglecting their experimental uncertainties.

\begin{table}[h]
\tbl{The values of the Fermi constant, of the masses of $h$, $Z$, $W$, and of the total width of $Z$ from Ref.~\citen{PDG:2014}.}
{\label{The values of G_F, m_h, m_Z, m_W, and Gamma_Z according to the 2014 data of Particle Data Group}
\begin{tabular}{l}
\hline 
$G_F = 1.1663787(6) \times 10^{-5} ~{\rm GeV}^{-2}$ \\
$m_h = 125.7(4) ~{\rm GeV}$ \\
$m_Z = 91.1876(21) ~{\rm GeV}$ \\
$m_W = 80.385(15) ~{\rm GeV}$ \\
$\Gamma_Z = 2.4952(23) ~{\rm GeV}$ \\
\hline 
\end{tabular}}
\end{table}

First, we show the SM distribution $\frac{1}{\Gamma} \frac{d^2 \Gamma}{d a_1 d a_2}$ for any decay $h \to Z_1^* Z_2^* \to f_1 \bar{f}_1 f_2 \bar{f}_2$ with $f_1$ different from $f_2$ (see Fig.~\ref{Plots of the distributions of a1 and a2}a) and that for any decay $h \to Z_1^* Z_2^* \to 4 l$ where $l$ stands for $e$, $\mu$, or $\tau$ (see Fig.~\ref{Plots of the distributions of a1 and a2}b). We see peaks at $\sqrt{a_1} = m_Z$ or $\sqrt{a_2} = m_Z$ and a flat surface outside the peaks for either dependence. For the decay into non-identical fermions the SM values of $\frac{1}{\Gamma} \frac{d^2 \Gamma}{d a_1 d a_2}$ on the peaks are about 120 times greater than the values on the ``plateau'' (the square $\sqrt{a_1}, \sqrt{a_2} \lesssim 50$ GeV). However, for the decay into identical leptons this ratio varies from 3 to 55 if we take $\sqrt{a_1} = m_Z$, $\sqrt{a_2} = \frac{1}{2} (m_h - m_Z)$ as the indicative point on the peak and on the plateau we consider the points on the line $\sqrt{a_1} = \sqrt{a_2}$ from $\sqrt{a_1} = 1$ GeV to $\sqrt{a_1} = 59$ GeV. Moreover, the SM probability that in a decay $h \to Z_1^* Z_2^* \to f_1 \bar{f}_1 f_2 \bar{f}_2$ either $Z$ boson has an invariant mass less than 50 GeV is
\begin{align}
\frac{1}{\Gamma_{f_1 \neq f_2} |_{SM}} \int \limits_0^{(50 ~{\rm GeV})^2} d a_1 \int \limits_0^{(50 ~{\rm GeV})^2} d a_2 \left. \frac{d^2 \Gamma_{f_1 \neq f_2}}{d a_1 d a_2} \right |_{SM} \approx 2.4 \%
\end{align}
while the corresponding probability for the decay $h \to Z_1^* Z_2^* \to 4 l$ is much higher, of about 21\%.

%

Figure~\ref{Plots of the distributions of a, theta, and phi} shows the distributions $\frac{1}{\Gamma} \frac{d \Gamma}{d a}$, $\frac{1}{\sin \theta} \frac{1}{\Gamma} \frac{d \Gamma}{d \theta}$, and $\frac{1}{\Gamma} \frac{d \Gamma}{d \phi}$ for the decay to non-identical leptons and the decay to identical ones. The definitions and explicit formulas for the differential widths $\frac{d \Gamma}{d a}$ and $\frac{d \Gamma}{d \theta}$ are given in \ref{Appendix: The definitions and explicit formulas for d Gamma over d a and d Gamma over d theta} (see Eqs.~(\ref{a definition of 1 over Gamma d Gamma over d a}), (\ref{an explicit formula for d Gamma over d a for any decay X-> Z_1^* Z_2^* -> f_1 antif_1 f_2 antif_2}), (\ref{a definition of 1 over Gamma d Gamma over d theta}), and (\ref{an explicit formula for d Gamma over d theta for any decay X-> Z_1^* Z_2^* -> f_1 antif_1 f_2 antif_2})).

\begin{figure}[h]

\begin{minipage}[h]{0.47 \linewidth}
\center{\includegraphics[width=0.85\linewidth]{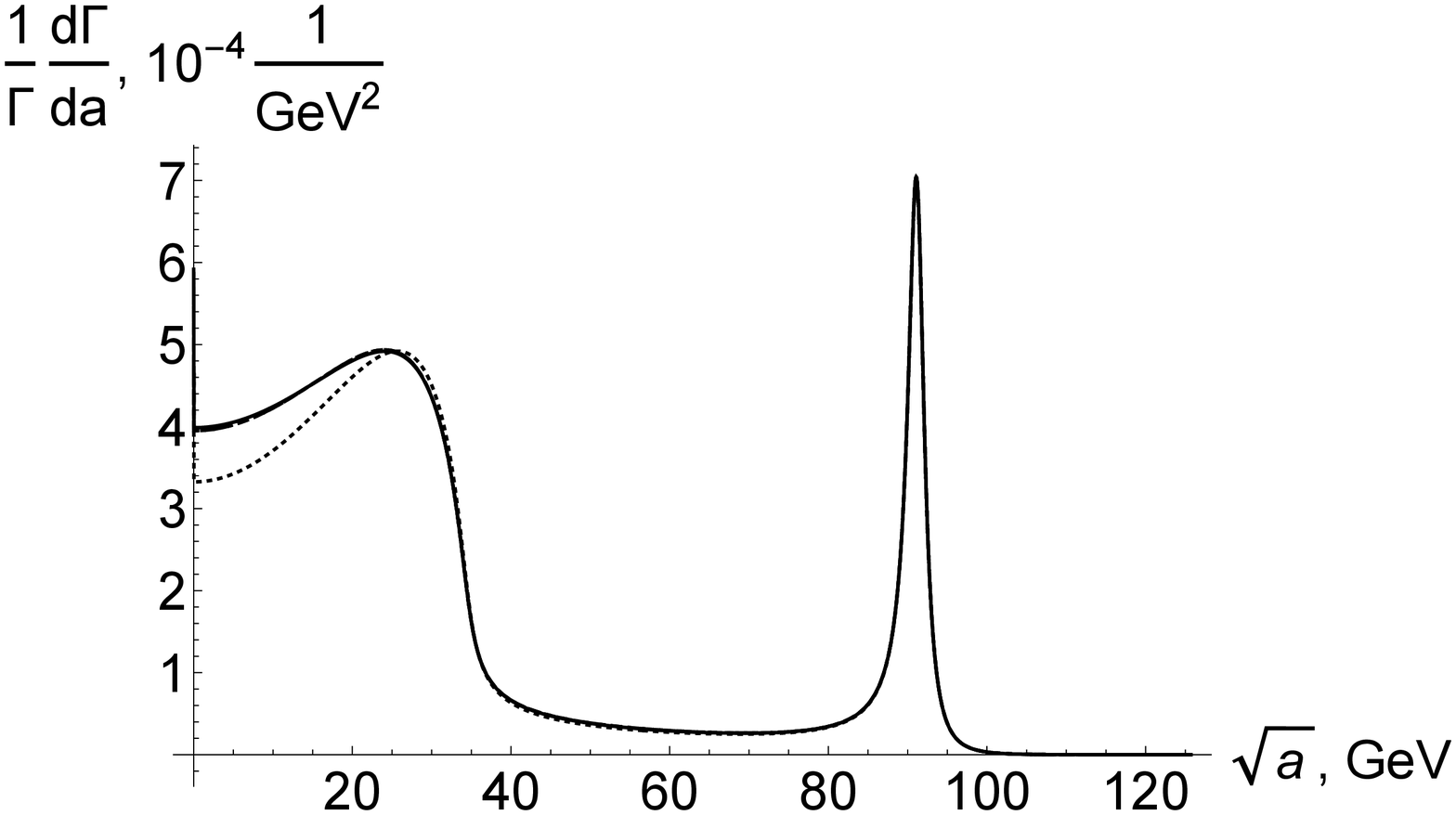}}
\end{minipage}
\hfill
\begin{minipage}[h]{0.47 \linewidth}
\center{\includegraphics[width=0.85\linewidth]{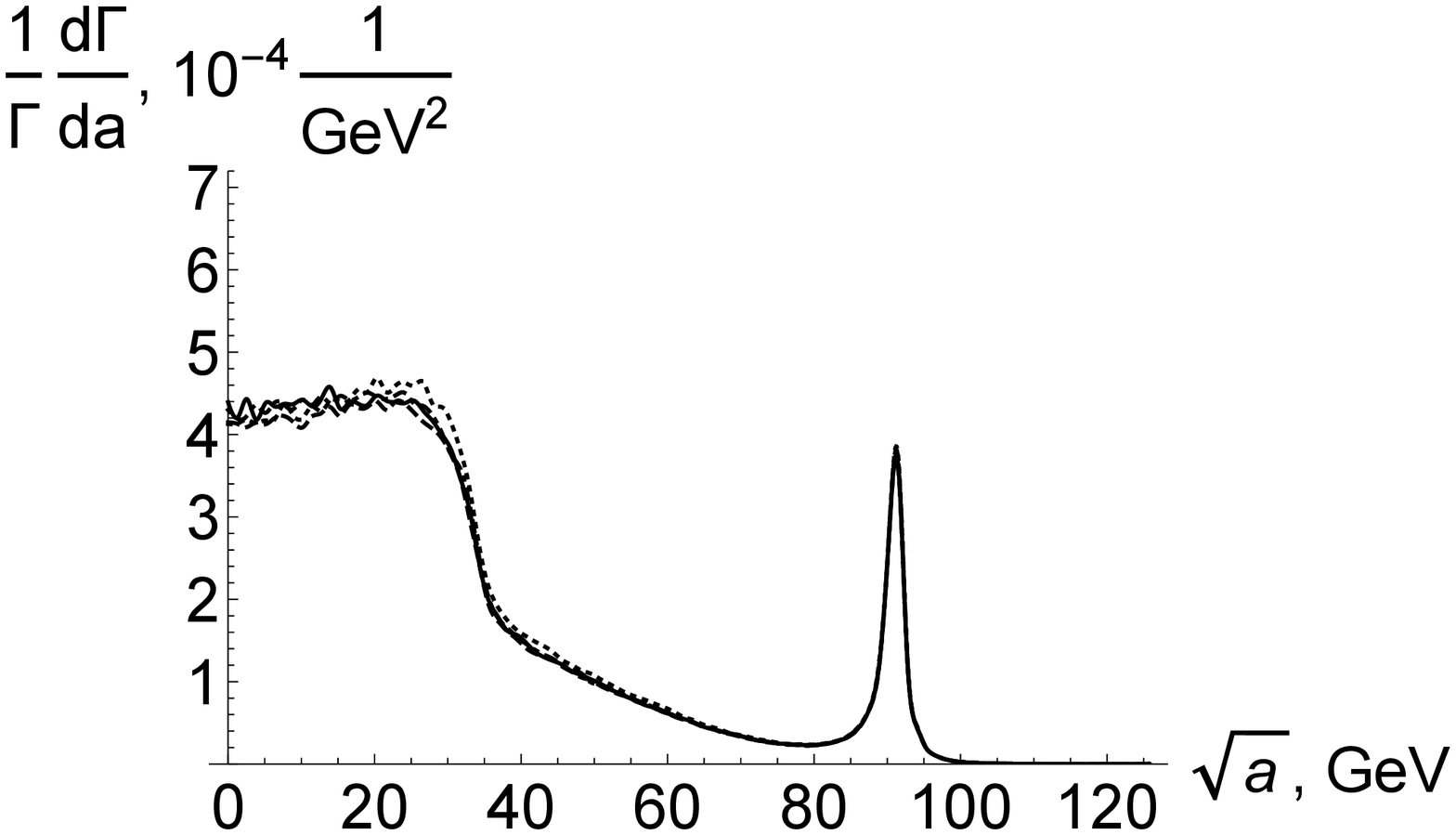}}
\end{minipage}
\vfill
\begin{minipage}[h]{0.47 \linewidth}
\center{\includegraphics[width=0.85\linewidth]{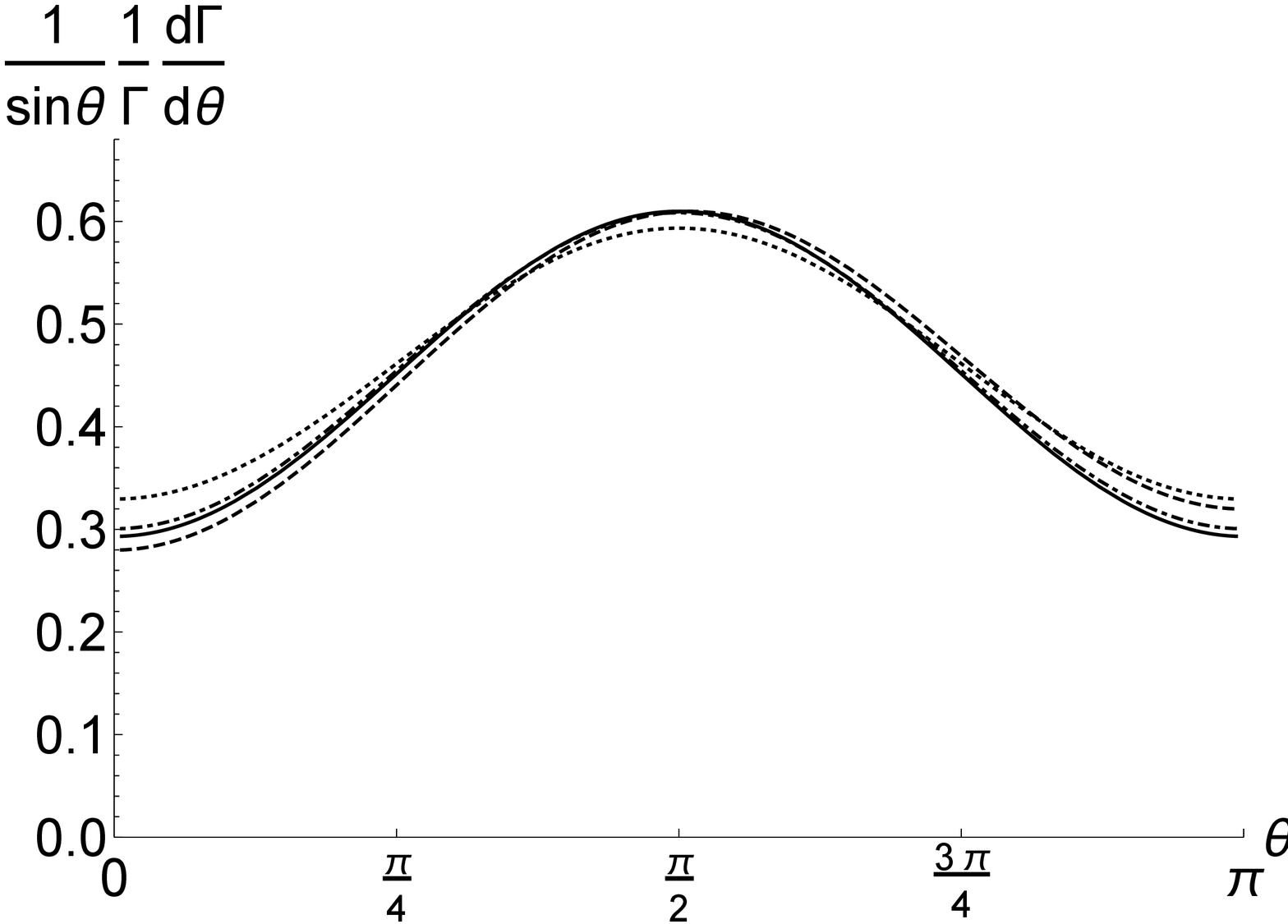}}
\end{minipage}
\hfill
\begin{minipage}[h]{0.47 \linewidth}
\center{\includegraphics[width=0.85\linewidth]{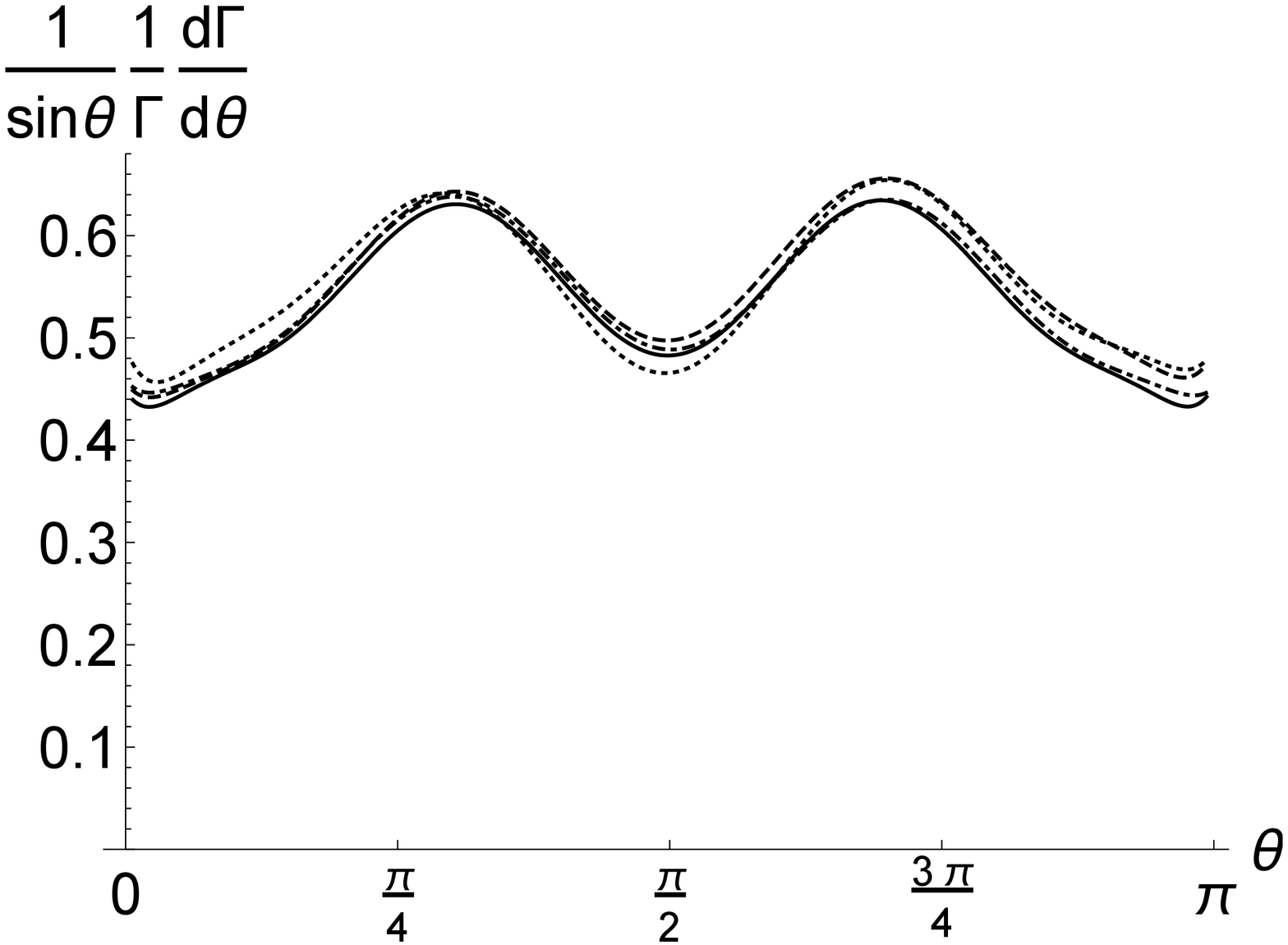}}
\end{minipage}
\vfill
\begin{minipage}[h]{0.47 \linewidth}
\center{\includegraphics[width=0.85\linewidth]{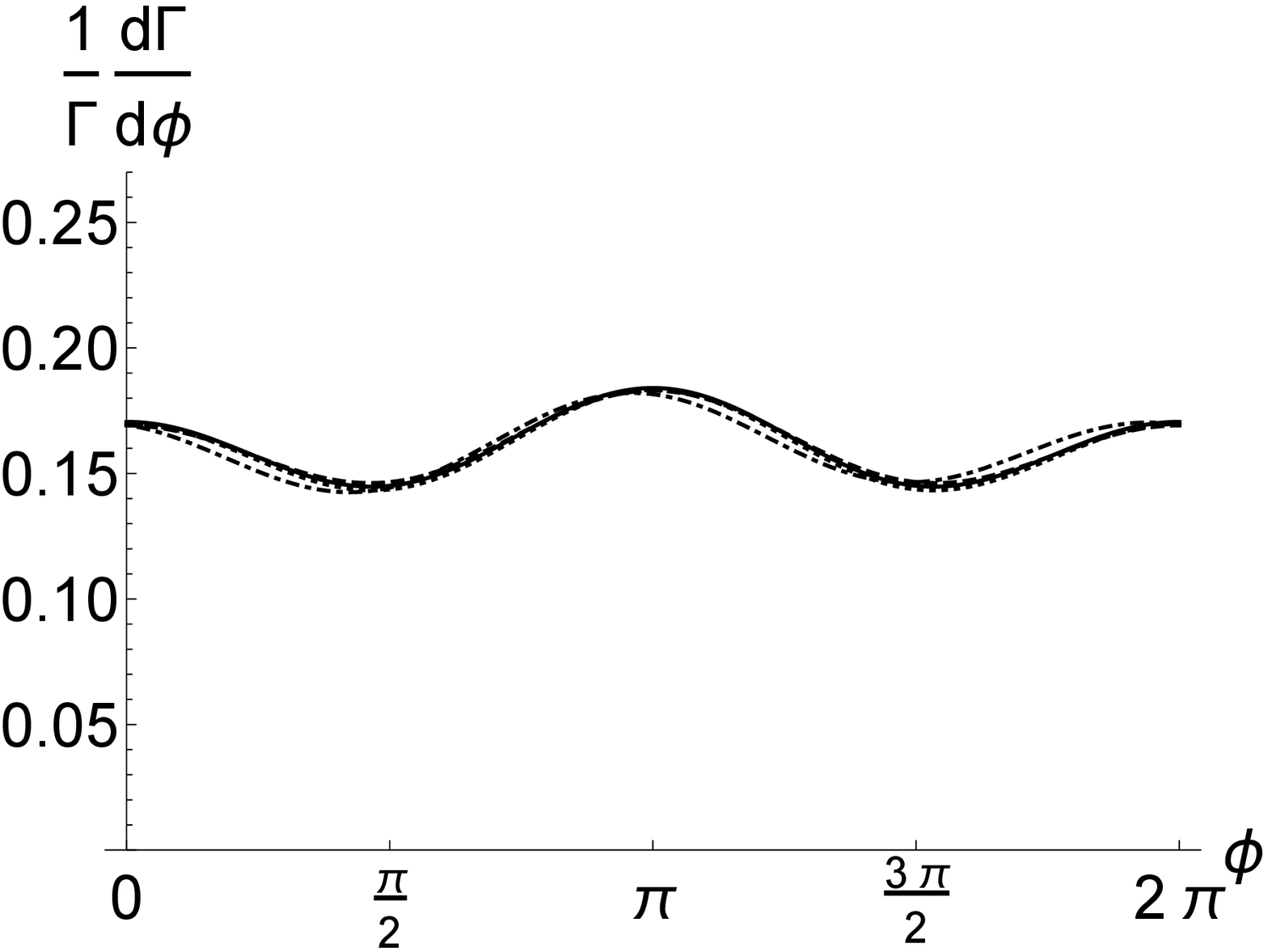} \\ a}
\end{minipage}
\hfill
\begin{minipage}[h]{0.47 \linewidth}
\center{\includegraphics[width=0.85\linewidth]{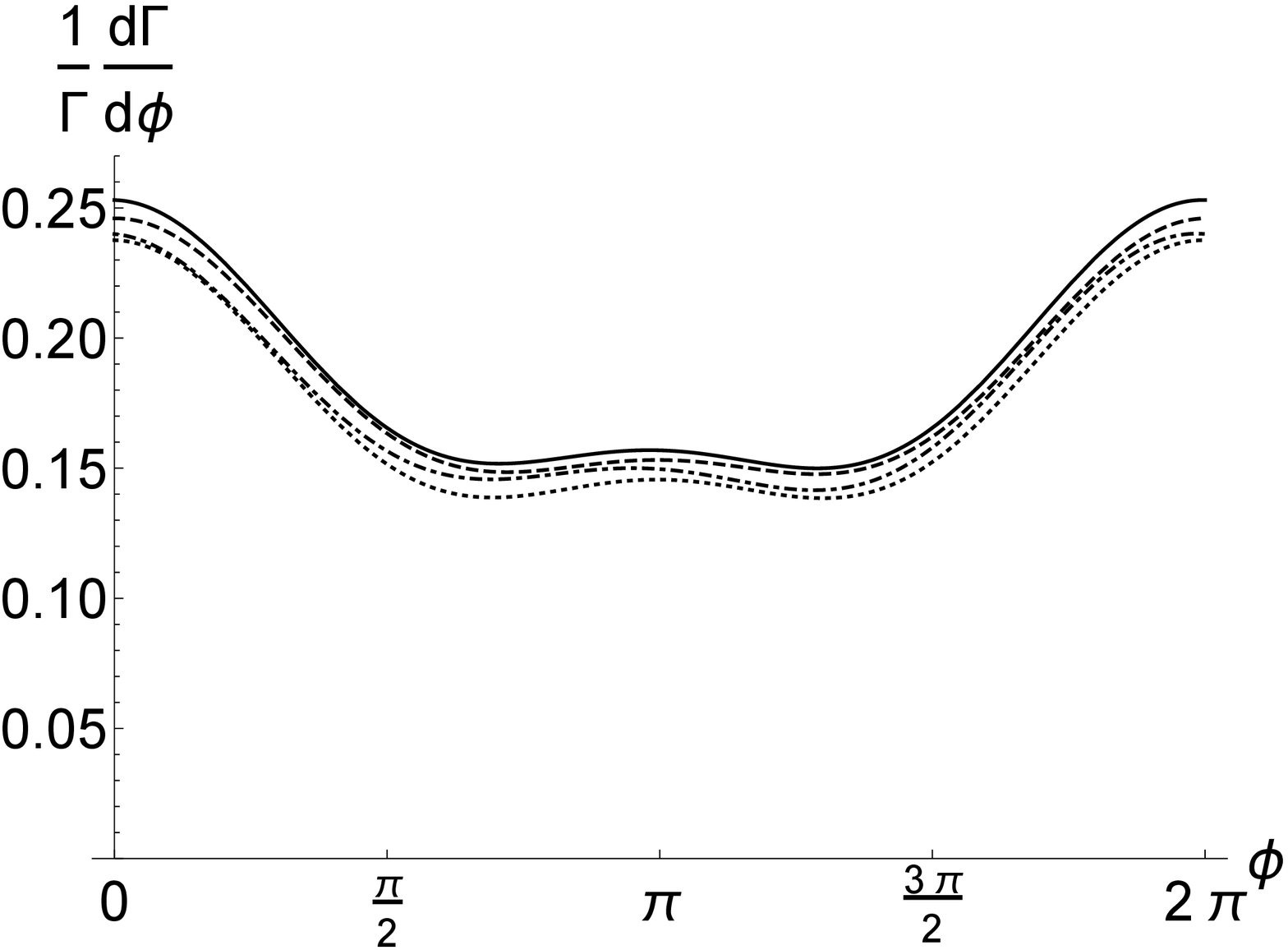} \\ b}
\end{minipage}

\caption{The distributions $\frac{1}{\Gamma}\frac{d \Gamma}{d a}$, $\frac{1}{\sin \theta} \frac{1}{\Gamma} \frac{d \Gamma}{d \theta}$, and $\frac{1}{\Gamma} \frac{d \Gamma}{d \phi}$ for the decays $h \to Z_1^* Z_2^* \to l_1^- l_1^+ l_2^- l_2^+$; $l_j = e, \mu, \tau$ in the cases $l_1 \neq l_2$ (a) and $l_1 = l_2$ (b). The solid, dashed, dot-dashed, and dotted lines correspond to sets (\ref{four sets of the possible values of a, b, c}) respectively.}
\label{Plots of the distributions of a, theta, and phi}
\end{figure}
%

The distributions in Fig.~\ref{Plots of the distributions of a, theta, and phi} are presented at the following four sets of values of the couplings $a_Z$, $b_Z$, and $c_Z$:
\begin{align}
\label{four sets of the possible values of a, b, c}
& |a_Z| = 1, \; b_Z = 0, \; c_Z = 0, & \notag \\
& a_Z = 1, \; b_Z = 0, \; c_Z = 0.5, & \notag \\
& a_Z = 1, \; b_Z = 0, \; c_Z = 0.5 \, i, & \notag \\
& a_Z = 1, \; b_Z = -0.5, \; c_Z = 0. &
\end{align}
In Ref.~\citen{Zagoskin:2016} sets (\ref{four sets of the possible values of a, b, c}) are shown to be consistent with the available LHC data and are chosen for an analysis of some observables sensitive to the $hZZ$ couplings.

The dependences in the upper plot of Fig.~\ref{Plots of the distributions of a, theta, and phi}a are calculated using Eq.~(A.2) from Ref.~\citen{Zagoskin:2016} and Eq.~(\ref{an explicit formula for d Gamma over d a for the decay into non-identical fermions}) from this paper. To obtain the lines shown in the two other plots of Fig.~\ref{Plots of the distributions of a, theta, and phi}a, we first integrate Eq.~(\ref{the fully differential width for the decays with f_1 different from f_2}) with a MC method and obtain four sets of dots. Then we fit each set by means of the method of least squares. In order not to clutter the plots, we show only the fitting lines and do not present the dots.

To derive the distributions $\frac{1}{\Gamma}\frac{d \Gamma}{d a}$, $\frac{1}{\sin \theta} \frac{1}{\Gamma} \frac{d \Gamma}{d \theta}$, and $\frac{1}{\Gamma} \frac{d \Gamma}{d \phi}$ for the decay into identical leptons, we integrate Eq.~(\ref{The fully differential width}) with a MC method and obtain sets of dots. The lines in the upper plot of Fig.~\ref{Plots of the distributions of a, theta, and phi}b consist of cubic parabolas joining the neighboring dots, since we have not been able to properly fit the dots of this plot with the method of least squares. The lines in the two other plots of Fig.~\ref{Plots of the distributions of a, theta, and phi}b are least-squares fits to the corresponding dots. As in  Fig.~\ref{Plots of the distributions of a, theta, and phi}a, the dots are not shown to avoid cluttering of the plots.

The relative uncertainties of the dots used for plotting the dependences in Fig.~\ref{Plots of the distributions of a, theta, and phi} are estimated during the MC integration. For any of the plotted distributions, these uncertainties turned out to be virtually the same for each dot and each set (\ref{four sets of the possible values of a, b, c}). Thus, they depend only on what distribution we consider. One standard deviation of a fitting line has been estimated using Eq.~(10) from Ref.~\citen{Richter:1995}. The uncertainties and one standard deviations for the distributions of the decays into non-identical or identical leptons are presented in Table~\ref{tab:uncertainties}. The estimates shown in Table~\ref{tab:uncertainties} do not account for the uncertainties of the constants listed in Table~\ref{The values of G_F, m_h, m_Z, m_W, and Gamma_Z according to the 2014 data of Particle Data Group}.

\begin{table}[h]
\tbl{The relative uncertainties $\delta_d$ of the dots and the standard deviations $\sigma_f$ of the fitting lines for some distributions of the decay $h \to Z_1^* Z_2^* \to l_1^- l_1^+ l_2^- l_2^+$ ($l_j = e, \mu, \tau$).}
{\label{tab:uncertainties}
\begin{tabular}{c | c  c c  c}
\hline 
Distribution & \multicolumn{2}{c}{non-identical leptons} & \multicolumn{2}{c}{identical leptons}  \\ \cline{2-3} \cline{4-5}
 & $\delta_d$  & $\sigma_f$  & $\delta_d$  & $\sigma_f$ \\
\hline
$\frac{1}{\Gamma}\frac{d \Gamma}{d a}$  & --  & -- & 1.8 \%  & -- \\
$\frac{1}{\sin \theta} \frac{1}{\Gamma} \frac{d \Gamma}{d \theta}$  & 2 \% & $1.2 \cdot 10^{-3}$ &  1.6 \% & $2.4 \cdot 10^{-3}$ \\
$\frac{1}{\Gamma} \frac{d \Gamma}{d \phi}$  & 2 \% & $5 \cdot 10^{-4}$  & 2 \% &  $7 \cdot 10^{-4}$ \\
\hline 
\end{tabular}}
\end{table}

We note that according to Fig.~3 in Ref.~\citen{Heinemeyer:2013}, the distinctions between the SM distributions $\frac{1}{\sin \theta} \frac{1}{\Gamma} \frac{d \Gamma}{d \theta}$ and $\frac{1}{\Gamma} \frac{d \Gamma}{d \phi}$ for the decay into non-identical leptons and those for the decay into identical ones are not as significant as these distinctions according to Fig.~\ref{Plots of the distributions of a, theta, and phi} in the present article. There can be a few sources of the differences with Fig.~3 in Ref.~\citen{Heinemeyer:2013}:

i) we consider the tree-level decays $h \to Z_1^* Z_2^* \to l_1^- l_1^+ l_2^- l_2^+$ while the dependences in Fig.~3 of Ref.~\citen{Heinemeyer:2013} are calculated at next-to-leading order (NLO) accuracy;

ii) we have numerically integrated Eq.~(8) from Ref.~\citen{Zagoskin:2016} and Eqs.~(\ref{the fully differential width for the decays with f_1 different from f_2}) and (\ref{The fully differential width}) from the present article, 
while MC integration with {\sc PROPHECY4f} was used in Ref.~\citen{Heinemeyer:2013};

iii) our definitions of the $Z$ boson couplings to fermions $a_f$ and $v_f$ and the asymmetry parameter $A_f$ are given in \ref{Appendix: A formula for the fully differential width of the decay into identical fermions}. These definitions yield $a_l = - 0.5$, $v_l = - 0.054$, and $A_l = 0.214$ ($l = e, \mu, \tau$). However, experimental values of these parameters are different. For instance, for the electron $a_e^{exp} = - 0.50123$, $v_e^{exp} = - 0.03783$, and $A_e^{exp} = 0.1515$ (see Ref.~\citen{PDG:2014}). The difference in $a_e$, $v_e$, and $A_e$ causes a certain distinction in the shapes of the distributions $\frac{1}{\sin \theta} \frac{1}{\Gamma} \frac{d \Gamma}{d \theta}$ and $\frac{1}{\Gamma} \frac{d \Gamma}{d \phi}$;  

iv) in the present article non-histrogram distributions are plotted.

The dependences plotted in Fig.~\ref{Plots of the distributions of a, theta, and phi} almost coincide at all four sets (\ref{four sets of the possible values of a, b, c}). For this reason, we can get significant constraints on $a_Z$, $b_Z$, and $c_Z$ via measurement of the distributions $\frac{1}{\Gamma} \frac{d \Gamma}{d a}$, $\frac{1}{\sin \theta} \frac{1}{\Gamma} \frac{d \Gamma}{d \theta}$, and $\frac{1}{\Gamma} \frac{d \Gamma}{d \phi}$ only if these distributions are measured at very high precision. That is why in order to constrain the $hZZ$ couplings, we should try to define observables sensitive to these couplings, like it is done in Ref.~\citen{Zagoskin:2016} for decay (\ref{X-> Z_1^* Z_2^* -> f_1 antif_1 f_2 antif_2, f_1 neq f_2}).

The distinctions between the distributions $\frac{1}{\Gamma} \frac{d \Gamma}{d a}$ for the decay into non-identical leptons (Fig.~\ref{Plots of the distributions of a, theta, and phi}a) and those for identical leptons (Fig.~\ref{Plots of the distributions of a, theta, and phi}b) are due to greater values of the SM distribution $\frac{1}{\Gamma} \frac{d^2 \Gamma}{d a_1 d a_2}$ on the plateau for the decay $h \to Z_1^* Z_2^* \to 4 l$ and smaller values of this distribution at the peaks $\sqrt{a_1} = m_Z$ and $\sqrt{a_2} = m_Z$ (see Fig.~\ref{Plots of the distributions of a1 and a2}). However, these distinctions are insubstantial.

The dissimilarity between the functions $\frac{1}{\sin \theta} \frac{1}{\Gamma} \frac{d \Gamma}{d \theta}$ and $\frac{1}{\Gamma} \frac{d \Gamma}{d \phi}$ in Figs.~\ref{Plots of the distributions of a, theta, and phi}a and \ref{Plots of the distributions of a, theta, and phi}b is much more appreciable. The global maximum of $\frac{1}{\sin \theta} \frac{1}{\Gamma} \frac{d \Gamma}{d \theta}$ at $\theta = \pi / 2$ in Fig.~\ref{Plots of the distributions of a, theta, and phi}a becomes a local minimum in Fig.~\ref{Plots of the distributions of a, theta, and phi}b, and the values near the points $\theta = 0$ and $\theta = \pi$ increase. Analogous distinctions take place between the dependences of $\frac{1}{\Gamma} \frac{d \Gamma}{d \phi}$ in Figs.~\ref{Plots of the distributions of a, theta, and phi}a and \ref{Plots of the distributions of a, theta, and phi}b.


\section{Comparison with experimental data}
\label{Section: Comparison with experimental data}

\subsection{ATLAS and CMS results}

In Ref.~\citen{Aad:2015} the ATLAS collaboration presents experimental distributions of the decay $h \to Z_1^* Z_2^* \to 4 \ell$ and corresponding distributions derived with MC simulations in the SM. We take the same kinematic limitations and the bin widths as ATLAS and use Eqs.~(\ref{The fully differential width}) and (\ref{the fully differential width for the decays with f_1 different from f_2}) to derive the SM histogram distributions of the decay $h \to Z_1^* Z_2^* \to 4 \ell$ which appear in Ref.~\citen{Aad:2015}. Comparison of our distributions with the ATLAS experimental and theoretical ones will determine the usefulness of Eq.~(\ref{The fully differential width}).

CMS has shown experimental distributions for the decay $h \to VV \to 4 \ell$ ($VV = ZZ$, $Z \gamma$, $\gamma \gamma$) and corresponding MC simulations in the SM in Ref.~\citen{Khachatryan:2015}. Taking the same kinematic limitations and the same bin widths as CMS, we integrate Eqs.~(\ref{The fully differential width}) and (\ref{the fully differential width for the decays with f_1 different from f_2}) in the SM to obtain distributions for the decay $h \to Z_1^* Z_2^* \to 4 \ell$.

We introduce the four following variables: $m_{12}$ ($m_{34}$) is the invariant mass of the $Z$ boson which is produced in a decay $h \to Z_1^* Z_2^* \to 4 \ell$ and whose mass is closest to (most distant from) $m_Z$, $\theta_1^{\prime}$ ($\theta_2^{\prime}$) is the polar angle of the fermion whose parent $Z$ boson has the invariant mass closest to (most distant from) $m_Z$. From the definitions of $m_{12}$ and $m_{34}$ it follows that
\begin{align}
|m_{12} - m_Z| < |m_{34} - m_Z|.
\end{align}
However, since $m_h < 2 \, m_Z$, the quantity $m_{12}$ ($m_{34}$) can be equivalently defined as the invariant mass of the heaviest (lightest) $Z$ boson produced in a decay $h \to Z_1^* Z_2^* \to 4 \ell$ ($m_{12} > m_{34}$).

In Ref.~\citen{Aad:2015} ATLAS shows distributions of $m_{12}$, $m_{34}$, $\cos \theta_1^{\prime}$, and $\phi$ (a distribution of $\cos \theta_2^{\prime}$ is not presented). ATLAS selects events $h \to Z_1^* Z_2^* \to 4 \ell$ wherein
\begin{align}
\label{the ATLAS limitations on m12, m34, eta_e, and eta_mu}
& m_{12} \in (50 ~{\rm GeV}, 106 ~{\rm GeV}), \qquad m_{34} \in (12 ~{\rm GeV}, 115 ~{\rm GeV}), & \notag \\
& \eta_e \in (-2.47, 2.47), \qquad \eta_{\mu} \in (-2.7, 2.7). &
\end{align}
Here $\eta_e$ ($\eta_{\mu}$) is the pseudorapidity of the electron (muon):
\begin{align}
\eta_i (\theta_i) \equiv - \ln \tan \frac{\theta_i}{2}, \qquad i = e, \mu,
\end{align}
where $\theta_e$ ($\theta_{\mu}$) is the polar angle of the electron (muon).

CMS paper \cite{Khachatryan:2015} presents distributions of $m_{12}$, $m_{34}$, $\cos \theta_1^{\prime}$, $\cos \theta_2^{\prime}$, and $\phi$ for the decay $h \to VV \to 4 \ell$ with
\begin{align}
\label{the CMS limitations on m12, m34, eta_e, and eta_mu}
& m_{12} \in (40 ~{\rm GeV}, 120 ~{\rm GeV}), \qquad m_{34} \in (12 ~{\rm GeV}, 120 ~{\rm GeV}), & \notag \\
& \eta_e \in (-2.5, 2.5), \qquad \eta_{\mu} \in (-2.4, 2.4). &
\end{align}

Constraints (\ref{the ATLAS limitations on m12, m34, eta_e, and eta_mu}) and (\ref{the CMS limitations on m12, m34, eta_e, and eta_mu}) determine the fractions of decays selected by ATLAS or CMS in the corresponding decay modes. These fractions are given by the left-hand sides of Eqs.~(\ref{the percentages of decays selected by ATLAS or CMS}) and (\ref{the percentage of decays h to ZZ to ell selected by ATLAS or CMS}). We have calculated the corresponding percentages in the SM (see Table~\ref{The SM percentages of decays selected by the CMS and ATLAS collaborations, for various decay modes}).

\begin{table}[h]
\tbl{The SM percentages $P_{SM}$ of decays selected by the CMS and ATLAS collaborations (see Eqs.~(\ref{the ATLAS limitations on m12, m34, eta_e, and eta_mu}) and (\ref{the CMS limitations on m12, m34, eta_e, and eta_mu})), for various decay modes.}
{\label{The SM percentages of decays selected by the CMS and ATLAS collaborations, for various decay modes}
\begin{tabular}{c | c | c}
\hline 
Decay mode & \multicolumn{2}{c}{$P_{SM}$} \\ \cline{2-3}
 & CMS & ATLAS \\
\hline
$h \to Z_1^* Z_2^* \to 4 e$ & 84.6 \% & 75.6 \% \\
$h \to Z_1^* Z_2^* \to 4 \mu$ & 84.1 \% & 76.4 \% \\
$h \to Z_1^* Z_2^* \to 2 e 2 \mu$ & 86.5 \% & 85.1 \% \\
$h \to Z_1^* Z_2^* \to 4 \ell$ & 85.5 \% & 81.1 \% \\
\hline 
\end{tabular}}
\end{table}

\subsection{A discussion of plots}

Integrating Eq.~(\ref{an explicit formula for the fully differential distribution of the decay h to Z1 Z2 to 4 ell with the ATLAS or CMS limitations}) with a MC method, we derive some SM histogram distributions of the decay $h \to Z_1^* Z_2^* \to 4 \ell$ (see the blue lines in Figs.~\ref{The numbers of events h to ZZ to 4 ell in bins of m_12, m_34, cos theta_1 prime, cos theta_2 prime, and phi with the ATLAS limitations} and \ref{The numbers of events h to ZZ to 4 ell in bins of m_12, m_34, cos theta_1 prime, cos theta_2 prime, and phi with the CMS limitations}). The bin widths in Fig.~\ref{The numbers of events h to ZZ to 4 ell in bins of m_12, m_34, cos theta_1 prime, cos theta_2 prime, and phi with the ATLAS limitations} are taken from Ref.~\citen{Aad:2015} while those in Fig.~\ref{The numbers of events h to ZZ to 4 ell in bins of m_12, m_34, cos theta_1 prime, cos theta_2 prime, and phi with the CMS limitations} are taken from Ref.~\citen{Khachatryan:2015}.

\begin{figure}[h]

\begin{minipage}[h]{0.47 \linewidth}
\center{\includegraphics[scale=0.65]{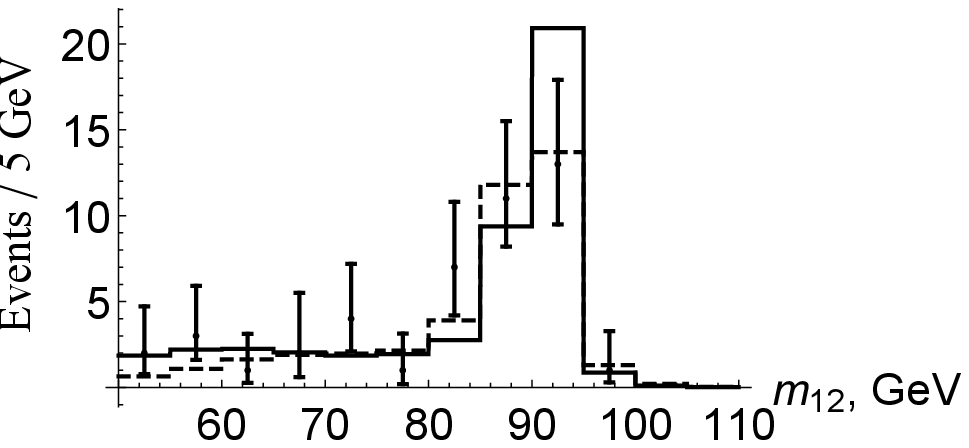}}
\end{minipage}
\hfill
\begin{minipage}[h]{0.47 \linewidth}
\center{\includegraphics[scale=0.65]{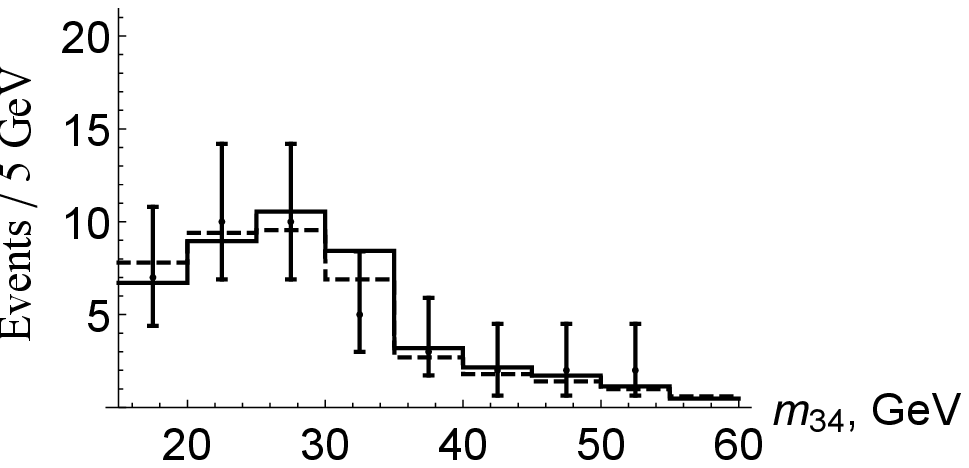}}
\end{minipage}
\vfill
\begin{minipage}[h]{0.47 \linewidth}
\center{\includegraphics[scale=0.65]{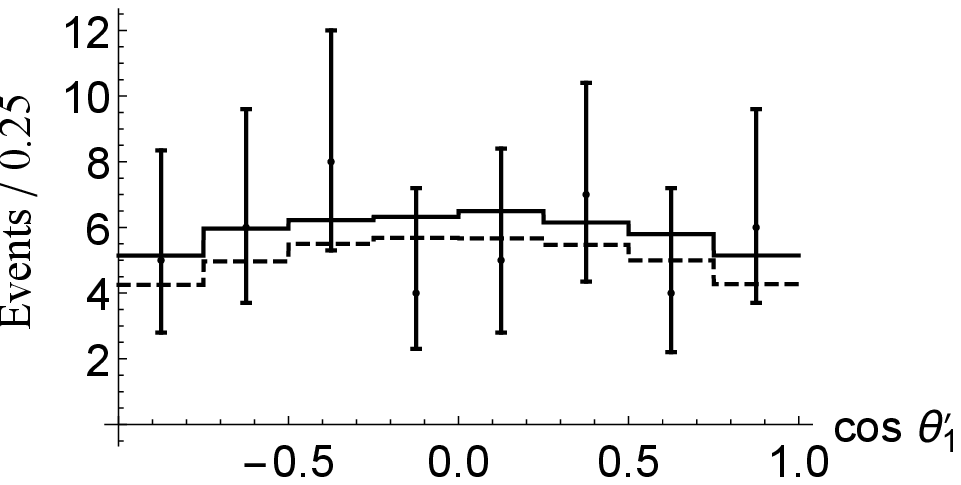}}
\end{minipage}
\hfill
\begin{minipage}[h]{0.47 \linewidth}
\center{\includegraphics[scale=0.65]{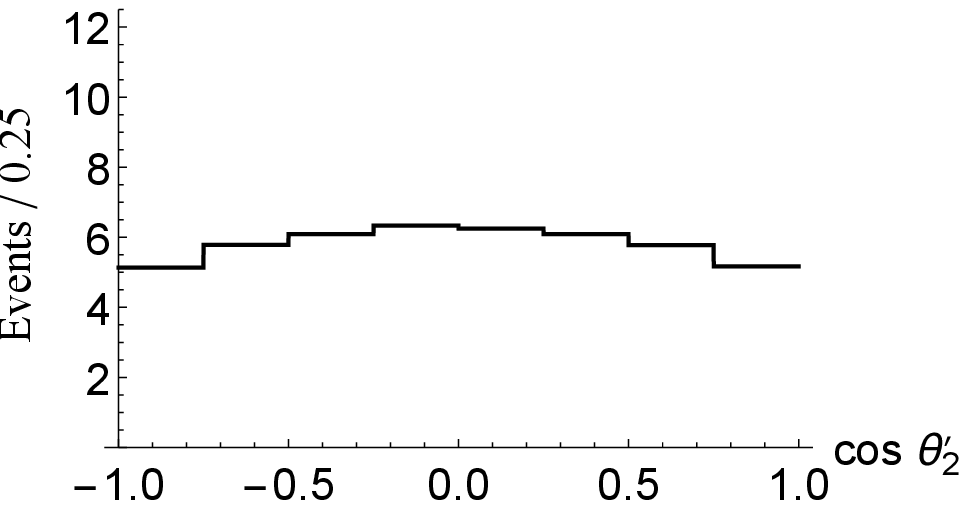}}
\end{minipage}
\vfill
\center{\includegraphics[scale=0.65]{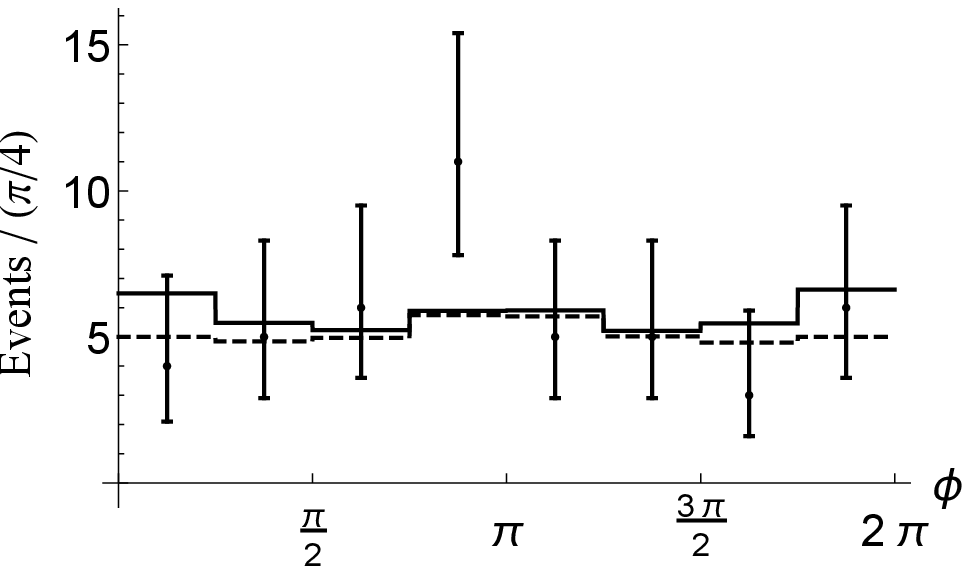}}

\caption{The numbers of events $h \to Z_1^* Z_2^* \to 4 \ell$ in bins of $m_{12}$, $m_{34}$, $\cos \theta_1^{\prime}$, $\cos \theta_2^{\prime}$, and $\phi$ according to our calculations in the SM (solid lines), the ATLAS (Ref.~\citen{Aad:2015}) MC simulations in the SM (dashed lines), and the ATLAS experimental data in Ref.~\citen{Aad:2015} (points with error bars). In our computations the total number of events $h \to Z_1^* Z_2^* \to 4 \ell$ is chosen to be 45. Both our calculations and the ATLAS MC simulations are carried out for ATLAS limitations (\ref{the ATLAS limitations on m12, m34, eta_e, and eta_mu}).}
\label{The numbers of events h to ZZ to 4 ell in bins of m_12, m_34, cos theta_1 prime, cos theta_2 prime, and phi with the ATLAS limitations}
\end{figure}
%

\begin{figure}[h]

\begin{minipage}[h]{0.47 \linewidth}
\center{\includegraphics[scale=0.65]{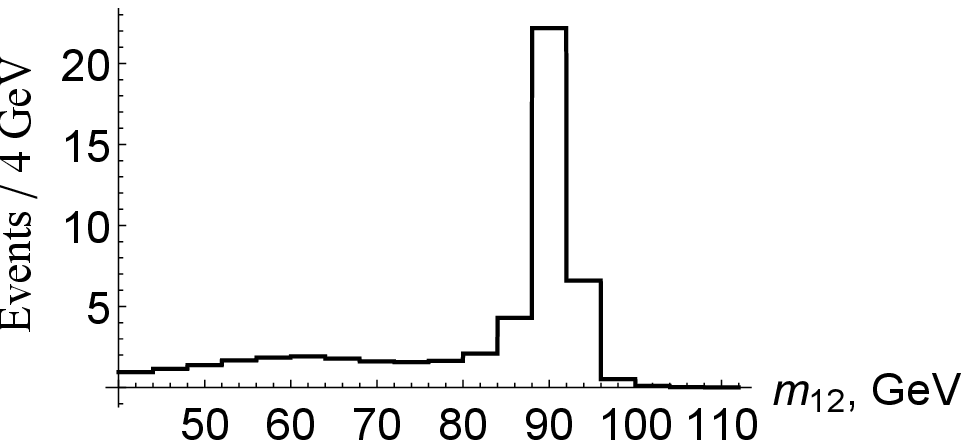}}
\end{minipage}
\hfill
\begin{minipage}[h]{0.47 \linewidth}
\center{\includegraphics[scale=0.65]{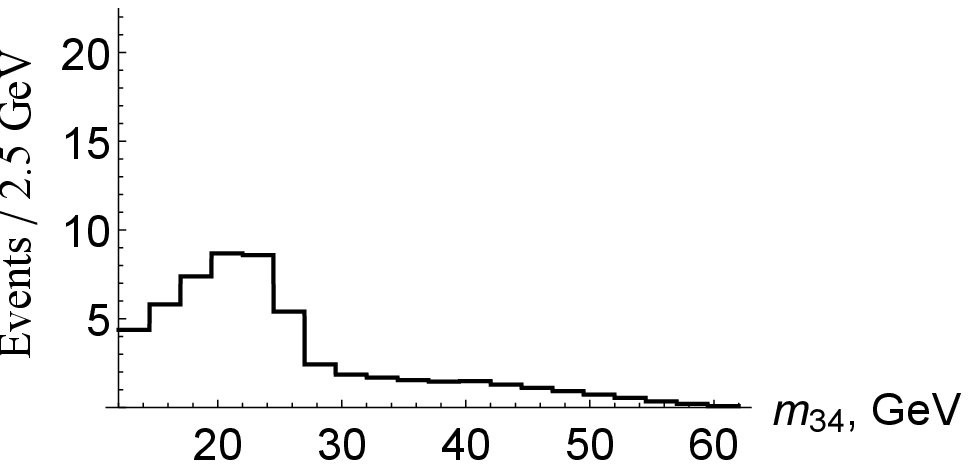}}
\end{minipage}
\vfill
\begin{minipage}[h]{0.47 \linewidth}
\center{\includegraphics[scale=0.65]{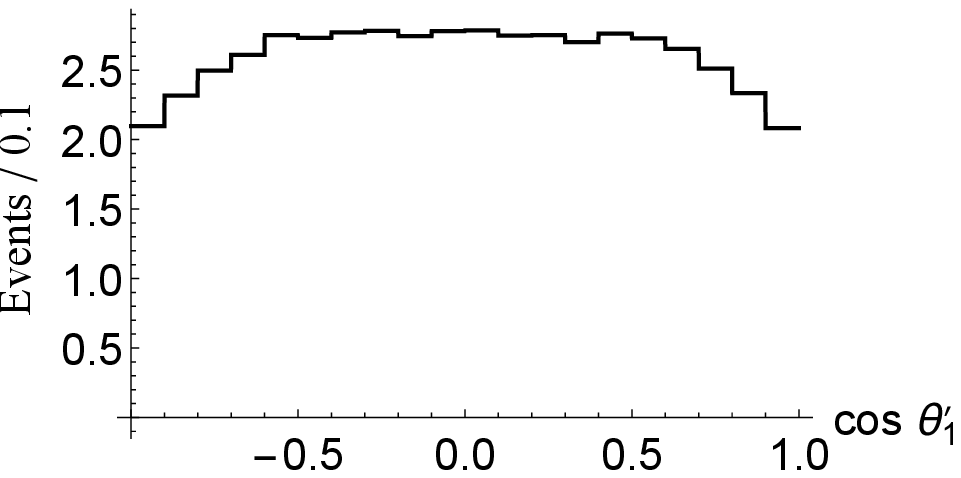}}
\end{minipage}
\hfill
\begin{minipage}[h]{0.47 \linewidth}
\center{\includegraphics[scale=0.65]{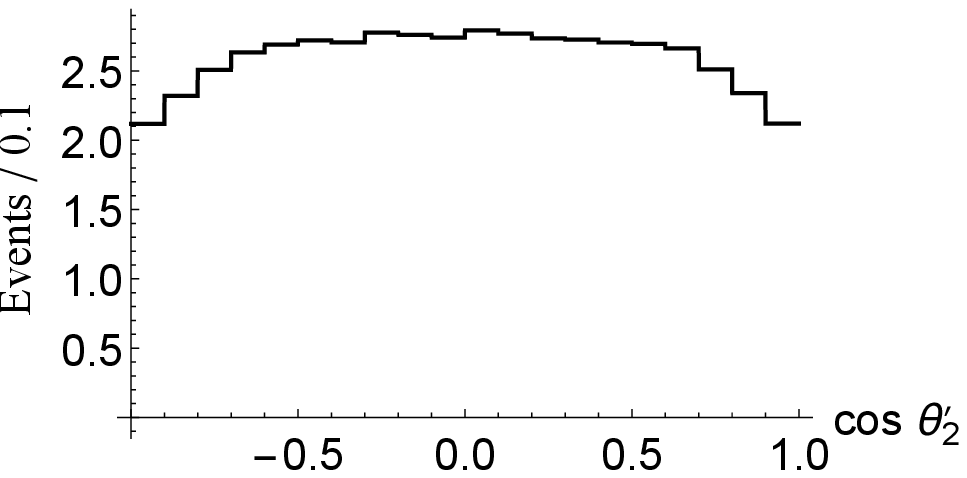}}
\end{minipage}
\vfill
\center{\includegraphics[scale=0.65]{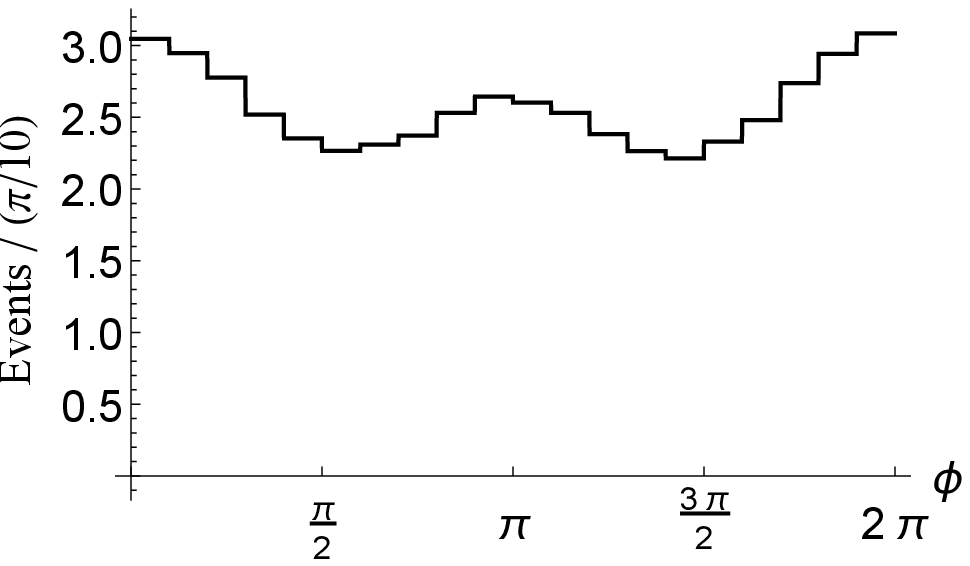}}

\caption{The numbers of events $h \to Z_1^* Z_2^* \to 4 \ell$ in bins of $m_{12}$, $m_{34}$, $\cos \theta_1^{\prime}$, $\cos \theta_2^{\prime}$, and $\phi$ according to our calculations in the SM. The total number of events $h \to Z_1^* Z_2^* \to 4 \ell$ is chosen to be 50. Our computations are performed for CMS limitations (\ref{the CMS limitations on m12, m34, eta_e, and eta_mu}).}
\label{The numbers of events h to ZZ to 4 ell in bins of m_12, m_34, cos theta_1 prime, cos theta_2 prime, and phi with the CMS limitations}
\end{figure}
%

ATLAS reports about 45 events $h \to Z_1^* Z_2^* \to 4 \ell$ with $m_{4 \ell} \in (115 \, {\rm GeV}, 130 \, {\rm GeV})$ ($m_{4 \ell}$ is the invariant mass of the 4 final leptons) in Ref.~\citen{Aad:2015} (see Table 3 there). For this reason, we have calculated our distributions shown in Fig.~\ref{The numbers of events h to ZZ to 4 ell in bins of m_12, m_34, cos theta_1 prime, cos theta_2 prime, and phi with the ATLAS limitations}, setting $N_{4 \ell}^{{\rm ATLAS}} = 45$ in Eq.~(\ref{an explicit formula for the fully differential distribution of the decay h to Z1 Z2 to 4 ell with the ATLAS or CMS limitations}).

It is of interest to sum up the numbers of events over all the bins for each plot in Fig.~\ref{The numbers of events h to ZZ to 4 ell in bins of m_12, m_34, cos theta_1 prime, cos theta_2 prime, and phi with the ATLAS limitations} (see Table~\ref{The sums over all the bins for each plot ATLAS}).

\begin{table}[h]
\tbl{The sums over all the bins for each plot in Fig.~\ref{The numbers of events h to ZZ to 4 ell in bins of m_12, m_34, cos theta_1 prime, cos theta_2 prime, and phi with the ATLAS limitations} ($\Sigma_{m_{12}}$, $\Sigma_{m_{34}}$,$\Sigma_{\cos \theta_1^{\prime}}$, $\Sigma_{\cos \theta_2^{\prime}}$, and $\Sigma_\phi$) for the ATLAS experimental data, for the ATLAS MC simulated distributions, and for our distributions.}
{\label{The sums over all the bins for each plot ATLAS}
\begin{tabular}{c | c | c | c}
\hline 
                                   & ATLAS exp. data & ATLAS MC simulated distributions & Our distributions \\ \hline
$\Sigma_{m_{12}}$   & 45                        & 40.31                                            & 46.16                  \\ \hline
$\Sigma_{m_{34}}$   & 41                        & 41.14                                            & 43.31                  \\ \hline
$\Sigma_{\cos \theta_1^{\prime}}$ & 45& 40.81                                            & 47.24                   \\ \hline
$\Sigma_{\cos \theta_2^{\prime}}$ &n/a& n/a                                                & 46.63                   \\ \hline
$\Sigma_\phi$            & 45                        & 41.09                                            & 46.30                   \\
\hline 
\end{tabular}}
\end{table}

The total number of the events in the ATLAS experimental distribution of $m_{34}$ is 41. That is why 4 events measured by ATLAS are not presented in this distribution. Therefore, in these events $m_{34} \in (12 \, {\rm GeV}, 15 \, {\rm GeV})$ (see ATLAS limitations (\ref{the ATLAS limitations on m12, m34, eta_e, and eta_mu}) and Fig.~\ref{The numbers of events h to ZZ to 4 ell in bins of m_12, m_34, cos theta_1 prime, cos theta_2 prime, and phi with the ATLAS limitations}). The bin sum 41.14 for the ATLAS simulated distribution of $m_{34}$ is notably closer to 41 than the bin sum 43.31 for our distribution of $m_{34}$.

For the ATLAS simulated distributions of $m_{12}$, $\cos \theta_1^{\prime}$, and $\cos \theta_2^{\prime}$ the bin sums are also close to 41. We take $N_{4 \ell}^{{\rm ATLAS}} = 45$ for all our distributions, and our bin sums $\Sigma_{m_{12}}$, $\Sigma_{\cos \theta_1^{\prime}}$, and $\Sigma_\phi$ are significantly closer to 45 than those for the ATLAS simulated distributions.

On the other hand, the ATLAS simulations take into account that for the 45 measured events $m_{4 \ell}$ varies from 115 GeV to 130 GeV while we use Eqs.~(\ref{The fully differential width}) and (\ref{the fully differential width for the decays with f_1 different from f_2}), which are derived for the case $m_{4 \ell} = m_h$.

Summarizing the comparison with the ATLAS results, we note that our distributions are derived by integration of analytical formulas obtained for $m_{4 \ell} = m_h$ and we have thoroughly chosen the total number of events. ATLAS has used MC simulations and has accounted for the fact that for the measured events $m_{4 \ell}$ varies from 115 GeV to 130 GeV. Both techniques have advantages and disadvantages, and therefore it is not surprising that the ATLAS simulated distributions and our distributions somewhat differ but are equally close to the ATLAS experimental distributions (see Fig.~\ref{The numbers of events h to ZZ to 4 ell in bins of m_12, m_34, cos theta_1 prime, cos theta_2 prime, and phi with the ATLAS limitations}). In addition, we present our distribution of $\cos \theta_2^{\prime}$.

In Ref.~\citen{Khachatryan:2015} CMS reports about 50 observed events $h \to VV \to 4 \ell$ with $m_{4 \ell} \in (105.6 \, {\rm GeV}, 140.6 \, {\rm GeV})$ (see Table~3 there). In view of this, in order to calculate distributions for the CMS limitations (\ref{the CMS limitations on m12, m34, eta_e, and eta_mu}), we choose $N_{4 \ell}^{{\rm CMS}} = 50$ in Eq.~(\ref{an explicit formula for the fully differential distribution of the decay h to Z1 Z2 to 4 ell with the ATLAS or CMS limitations}). The accuracy of our distributions shown in Fig.~\ref{The numbers of events h to ZZ to 4 ell in bins of m_12, m_34, cos theta_1 prime, cos theta_2 prime, and phi with the CMS limitations} can be characterized by the sums over all the bins for each plot (see Table~\ref{The sums over all the bins for each plot CMS}). The plots in Fig.~\ref{The numbers of events h to ZZ to 4 ell in bins of m_12, m_34, cos theta_1 prime, cos theta_2 prime, and phi with the CMS limitations} are smoother than those in Fig.~\ref{The numbers of events h to ZZ to 4 ell in bins of m_12, m_34, cos theta_1 prime, cos theta_2 prime, and phi with the ATLAS limitations} due to their smaller bin widths.

\begin{table}[h]
\tbl{The sums over all the bins for each plot in Fig.~\ref{The numbers of events h to ZZ to 4 ell in bins of m_12, m_34, cos theta_1 prime, cos theta_2 prime, and phi with the CMS limitations}.}
{\label{The sums over all the bins for each plot CMS}
\begin{tabular}{c | c}
\hline 
                                                               & Our distributions \\ \hline
$\Sigma_{m_{12}}$                               & 51.30                  \\ \hline
$\Sigma_{m_{34}}$                               & 55.91                  \\ \hline
$\Sigma_{\cos \theta_1^{\prime}}$     & 52.14                   \\ \hline
$\Sigma_{\cos \theta_2^{\prime}}$     & 52.03                   \\ \hline
$\Sigma_\phi$                                        & 51.34                   \\
\hline 
\end{tabular}}
\end{table}

\section{Conclusions}
\label{Section: Conclusions}

In this paper, we have considered the decay of a neutral particle $X$ with zero spin and arbitrary $CP$ parity into two off-mass-shell $Z$ bosons ($Z_1^*$ and $Z_2^*$) each of which decays to identical fermion-antifermion pairs ($f \bar{f}$): $X \to Z_1^* Z_2^* \to f \bar{f} f \bar{f}$. Analytical formulas for the fully differential width of the decay in question and for the fully differential width of the decay $h \to Z_1^* Z_2^* \to 4 \ell$ are derived (see Eqs.~(\ref{The fully differential width}) and (\ref{an explicit formula for the fully differential distribution of the decay h to Z1 Z2 to 4 ell})). Moreover, we present an exact formula for the differential width $\frac{d \Gamma} {d a}$ of a decay $X \to Z_1^* Z_2^* \to f_1 \bar{f}_1 f_2 \bar{f}_2$ with $f_1 \neq f_2$ (see Eq.~(\ref{an explicit formula for d Gamma over d a for the decay into non-identical fermions})).

Integrating Eq.~(\ref{The fully differential width}) with a MC method, we have obtained some non-histogram distributions for any decay $h \to Z_1^* Z_2^* \to l_1^- l_1^+ l_2^- l_2^+$ ($l_j = e, \mu, \tau$) with $l_1 = l_2$. These distributions are compared to those for the decay $h \to Z_1^* Z_2^* \to l_1^- l_1^+ l_2^- l_2^+$ with $l_1 \neq l_2$ (see Figs.~\ref{Plots of the distributions of a1 and a2} and \ref{Plots of the distributions of a, theta, and phi}). The comparison has revealed significant distinctions between the distributions for the case $l_1 = l_2$ and the corresponding ones for $l_1 \neq l_2$. However, in the SM some of these distinctions may be less noticeable, as Figure 3 in Ref.~\citen{Heinemeyer:2013} presents. The difference between the results of Ref.~\citen{Heinemeyer:2013} and our ones can arise due to several reasons discussed in Section~\ref{Section: Invariant mass and angular distributions}. The dependences shown in Fig.~\ref{Plots of the distributions of a, theta, and phi} are calculated at four possible sets (\ref{four sets of the possible values of a, b, c}) of values of the $hZZ$ couplings $a_Z$, $b_Z$, and $c_Z$. At all the four sets these distributions almost coincide. Therefore their measurement can yield notable constraints on $a_Z$, $b_Z$, and $c_Z$ only if the distributions are measured at very high precision.

In order to determine the usefulness of Eq.~(\ref{The fully differential width}), we have computed some SM histogram distributions of the decay $h \to Z_1^* Z_2^* \to 4 \ell$ by means of integration of Eq.~(\ref{an explicit formula for the fully differential distribution of the decay h to Z1 Z2 to 4 ell with the ATLAS or CMS limitations}). The distributions are calculated for ATLAS kinematical limitations (\ref{the ATLAS limitations on m12, m34, eta_e, and eta_mu}) and for CMS ones (\ref{the CMS limitations on m12, m34, eta_e, and eta_mu}).

We have compared our distributions with the ATLAS experimental ones and the ATLAS MC simulated ones (see Ref.~\citen{Aad:2015}). The way our distributions are derived is almost purely analytical --- its only numerical part is integration of Eq.~(\ref{an explicit formula for the fully differential distribution of the decay h to Z1 Z2 to 4 ell with the ATLAS or CMS limitations}). Besides, we have chosen the total number of events more accurately than ATLAS during its simulations. However, our calculation does not allow for the fact that the invariant mass of $4 \ell$ may differ from $m_h$ while this fact is taken into account in the ATLAS simulations. The pros and cons of our technique and the ATLAS simulations make our distributions and the ATLAS simulated ones somewhat different but equally close to the ATLAS experimental data.

We have also presented our distributions of $m_{12}$, $m_{34}$, $\cos \theta_1^\prime$, $\cos \theta_2^\prime$, and $\phi$ for the kinematic conditions specific for CMS.

In summary, various distributions of the decays $X \to Z_1^* Z_2^* \to f \bar{f} f \bar{f}$ or $h \to Z_1^* Z_2^* \to 4 \ell$ have been obtained with a rather simple integration of Eqs.~(\ref{The fully differential width}) and (\ref{an explicit formula for the fully differential distribution of the decay h to Z1 Z2 to 4 ell}) respectively. This way of calculation gives an alternative to more traditional MC simulation.

\section*{Acknowledgments}
This research was partially supported by the National Academy of Sciences of Ukraine (project TsO-1-4/2017) and the Ministry of Education and Science of Ukraine (projects no. 0115U000473 and 0117U004866).


\appendix

\section{The fully differential width of the decay $X \to Z_1^* Z_2^* \to f \bar{f} f \bar{f}$}
\label{Appendix: A formula for the fully differential width of the decay into identical fermions}

The fully differential width of decay (\ref{X-> Z_1^* Z_2^* -> f antif f antif}) is
\begin{align}
\label{The fully differential width}
\frac{d^5 \Gamma}{da_1 da_2 d \theta_1 d \theta_2 d \varphi} = & \frac{1}{4} \Biggl [\left. \frac{d^5 \Gamma_{f_1 \neq f_2}} {da_1 da_2 d \theta_1 d \theta_2 d \varphi} \right |_{f_1 = f_2 = f} + \frac{\sqrt{2} G_F^3 m_Z^8}{(4 \pi)^6 m_X^3} \frac{k \sqrt{\tilde{a}_1 \tilde{a}_2}}{D (\tilde{a}_1) D (\tilde{a}_2)} (a_f^2 + v_f^2)^2 \sin \theta_1 \sin \theta_2 & \notag \\
& \times \biggl \{ \sqrt{\tilde{a}_1 \tilde{a}_2} \Bigl \{ \left ((1 + \bar{\alpha}_3^2) (1 + \bar{\beta}_3^2) + 4 A_f^2 \bar{\alpha}_3 \bar{\beta}_3 \right) (|\tilde{A}_{\parallel}|^2 + |\tilde{A}_{\perp}|^2) + 4 (1 - \bar{\alpha}_3^2) (1 - \bar{\beta}_3^2) |\tilde{A}_0|^2 & \notag \\
& - 4 A_f \left (\bar{\alpha}_3 (1 + \bar{\beta}_3^2) + \bar{\beta}_3 (1 + \bar{\alpha}_3^2) \right) {\rm Re} (\tilde{A}_{\parallel}^* \tilde{A}_{\perp}) + 4 \sqrt{2} \Bigl ( (A_f^2 + \bar{\alpha}_3 \bar{\beta}_3) \bigl ({\rm Re} \eta_- {\rm Re} (\tilde{A}_0^* \tilde{A}_{\parallel}) & \notag \\
& + {\rm Im} \eta_- {\rm Im} (\tilde{A}_0^* \tilde{A}_{\perp}) \bigr) - A_f (\bar{\alpha}_3 + \bar{\beta}_3) \bigl ({\rm Re} \eta_- {\rm Re} (\tilde{A}_0^* \tilde{A}_{\perp}) + {\rm Im} \eta_- {\rm Im} (\tilde{A}_0^* \tilde{A}_{\parallel}) \bigr) \Bigr) & \notag \\
& + {\rm Re} \eta_-^2 (|\tilde{A}_{\parallel}|^2 - |\tilde{A}_{\perp}|^2) + 2 \, {\rm Im} \eta_-^2 {\rm Im} (\tilde{A}_{\parallel}^* \tilde{A}_{\perp}) \Bigl \} - \frac{\sqrt{a_1 a_2}}{D (a_1) D (a_2)} & \notag \\
& \times {\rm Re} \Bigl\{ (a_1 - m_Z^2 + i m_Z \Gamma_Z) (a_2 - m_Z^2 + i m_Z \Gamma_Z) (\tilde{a}_1 - m_Z^2 - i m_Z \Gamma_Z) (\tilde{a}_2 - m_Z^2 - i m_Z \Gamma_Z) & \notag \\
& \times \Bigl ( \bigl ( (r_{\alpha \beta} + \frac{1}{r_{\alpha \beta}}) ({\rm Re} \eta_- \tilde{A}_{\parallel} - i {\rm Im} \eta_- \tilde{A}_{\perp}) + 2 \sqrt{2 (1 - \bar{\alpha}_3^2) (1 - \bar{\beta}_3^2)} \tilde{A}_0 \bigr) & \notag \\
& \times \bigl ( ((1 + A_f^2) \cos \phi (1 + \cos \theta_1 \cos \theta_2) + i \cdot 2 A_f \sin \phi (\cos \theta_1 + \cos \theta_2)) A_{\parallel}^* & \notag \\
& - (2 A_f \cos \phi (\cos \theta_1 + \cos \theta_2) + i (1 + A_f^2) \sin \phi (1 + \cos \theta_1 \cos \theta_2)) A_{\perp}^* & \notag \\
& + \sqrt{2} (1 + A_f^2) \sin \theta_1 \sin \theta_2 A_0^* \bigr) + (r_{\alpha \beta} - \frac{1}{r_{\alpha \beta}}) (i {\rm Im} \eta_- \tilde{A}_{\parallel} - {\rm Re} \eta_- \tilde{A}_{\perp}) & \notag \\
& \times \bigl ((2 A_f \cos \phi (1 + \cos \theta_1 \cos \theta_2) + i (1 + A_f^2) \sin \phi (\cos \theta_1 + \cos \theta_2)) A_{\parallel}^* & \notag \\
& - ((1 + A_f^2) \cos \phi (\cos \theta_1 + \cos \theta_2) + i \cdot 2 A_f \sin \phi (1 + \cos \theta_1 \cos \theta_2)) A_{\perp}^* & \notag \\
& + \sqrt{2} \cdot 2 A_f \sin \theta_1 \sin \theta_2 A_0^* \bigr) \Bigr ) \Bigr\} \biggr \} \Biggr], &
\end{align}
where
\begin{align}
\label{the fully differential width for the decays with f_1 different from f_2}
\frac{d^5 \Gamma_{f_1 \neq f_2}} {da_1 da_2 d \theta_1 d \theta_2 d \varphi} = & \frac {\sqrt{2} G_F^3 m_Z^8} {(4 \pi)^6 m_X^3} (a_{f_1}^2 + v_{f_1}^2) (a_{f_2}^2 + v_{f_2}^2) \frac{k a_1 a_2}
{ D(a_1) D(a_2)}  & \notag \\
& \times \sin \theta_1 \sin \theta_2 [(|A_{\parallel}|^2 + |A_{\perp}|^2) \left ((1 + \cos^2 \theta_1) (1 + \cos^2 \theta_2)
 + 4 A_{f_1} A_{f_2} \cos \theta_1 \cos \theta_2 \right)  & \notag \\
& + 4 |A_0|^2 \sin^2 \theta_1 \sin^2 \theta_2 - 4 \, {\rm Re}(A_{\parallel}^* A_{\perp}) (A_{f_1} \cos \theta_1 (1 + \cos^2 \theta_2) + A_{f_2} \cos \theta_2 (1 + \cos^2 \theta_1))  & \notag \\
& + 4 \sqrt{2} \sin \theta_1 \sin \theta_2 (({\rm  Re}(A_0^* A_{\parallel}) \cos \phi - {\rm Im}(A_0^* A_{\perp}) \sin \phi) (A_{f_1} A_{f_2} + \cos \theta_1 \cos \theta_2)  & \notag \\
& - ({\rm  Re}(A_0^* A_{\perp}) \cos \phi - {\rm Im}(A_0^* A_{\parallel}) \sin \phi) (A_{f_1} \cos \theta_2 + A_{f_2} \cos \theta_1))  & \notag \\
& + \sin^2 \theta_1 \sin^2 \theta_2 ((|A_{\parallel}|^2 - |A_{\perp}|^2) \cos 2 \phi - 2 \, {\rm Im}(A_{\parallel}^* A_{\perp}) \sin 2 \phi) ] &
\end{align}
is the fully differential width of decay (\ref{X-> Z_1^* Z_2^* -> f_1 antif_1 f_2 antif_2, f_1 neq f_2}) (see Eq.~(5) in Ref.~\citen{Zagoskin:2016}), $a_f$ is the weak isospin projection of the fermion $f$, $v_f \equiv a_f - 2 \frac{q_f}{e} \sin^2 \theta_W$, $q_f$ is the electric charge of $f$, $e$ is the electric charge of the positron, $\theta_W$ is the weak mixing angle, $D (x) \equiv (x - m_Z^2)^2 + (m_Z \Gamma_Z)^2$,
\begin{align}
A_{\pm} \equiv \frac{A_{X \rightarrow Z_1^* Z_2^*} (p_1, p_2, \pm 1, \pm 1)} {g_Z}, \qquad A_0 \equiv \frac{A_{X \rightarrow Z_1^* Z_2^*} (p_1, p_2, 0, 0)} {g_Z},
\end{align}
\begin{align}
A_{\parallel} \equiv \frac{\tilde{A}_+ + \tilde{A}_-}{\sqrt{2}} = \sqrt{2} a_Z (a_1, a_2), \qquad A_{\perp} \equiv \frac{\tilde{A}_+ - \tilde{A}_-}{\sqrt{2}} = \sqrt{2} \frac{k}{m_X^2} c_Z (a_1, a_2),
\end{align}
\begin{align}
\label{a definition of A_f}
A_f = \frac{2 a_f v_f}{a_f^2 + v_f^2},
\end{align}
$\bar{\alpha}_i \equiv \frac{\alpha_i}{|\boldsymbol \alpha|} ~(i = 1, 2, 3)$, $\boldsymbol \alpha$ is the momentum of the fermion $f_1$ in the center-of-momentum frame of the particles $f_1$ and $\bar{f}_2$,
\begin{align}
\boldsymbol \alpha = & \mathbf{e}_x \frac{\sqrt{a_1} (2 E_2^{\prime} + \sqrt{\tilde{a}_1}) \sin \theta_1 \cos \phi_1 + \sqrt{a_2} (2 E_1 + \sqrt{\tilde{a}_1}) \sin \theta_2 \cos \phi_2}{4 (E_1 + E_2^{\prime} + \sqrt{\tilde{a}_1})} & \notag \\
& + \mathbf{e}_y \frac{\sqrt{a_1} (2 E_2^{\prime} + \sqrt{\tilde{a}_1}) \sin \theta_1 \sin \phi_1 - \sqrt{a_2} (2 E_1 + \sqrt{\tilde{a}_1}) \sin \theta_2 \sin \phi_2}{4 (E_1 + E_2^{\prime} + \sqrt{\tilde{a}_1})} + \frac{\bar{\mathbf{p}}_1}{8 (E_1 + E_2^{\prime} + \sqrt{\tilde{a}_1})} & \notag \\
& \times \Bigl ((m_X^2 - a_1 - a_2) (\cos \theta_1 - \cos \theta_2) + k (1 - \cos \theta_1 \cos \theta_2) & \notag \\
& + \frac{\sqrt{\tilde{a}_1}}{m_X} (2 k + (m_X^2 + a_1 - a_2) \cos \theta_1 - (m_X^2 + a_2 - a_1) \cos \theta_2) \Bigr), &  \\
|\boldsymbol \alpha| = & \frac{\sqrt{\tilde{a}_1}}{2}, &
\end{align}
$\bar{\mathbf{p}}_1 \equiv \frac{\mathbf{p}_1}{|\mathbf{p}_1|}$, $\mathbf{e}_x$ and $\mathbf{e}_y$ are any unit and mutually orthogonal vectors such that $\mathbf{e}_x \times \mathbf{e}_y = \bar{\mathbf{p}}_1$,
\begin{align}
E_1 \equiv k_1^0 = \frac{m_X^2 + a_1 - a_2 + k \cos \theta_1}{4 m_X}, \qquad  E_1^{\prime} \equiv k_1^{\prime 0} = \frac{m_X^2 + a_1 - a_2 - k \cos \theta_1}{4 m_X} \notag \\
E_2 \equiv k_2^0 = \frac{m_X^2 + a_2 - a_1 + k \cos \theta_2}{4 m_X}, \qquad E_2^{\prime} \equiv k_2^{\prime 0} = \frac{m_X^2 + a_2 - a_1 - k \cos \theta_2}{4 m_X},
\end{align}
$\phi_1$ is the azimuthal angle of the $f_1$ momentum in the $Z_1^*$ rest frame formed by the vectors ($\mathbf{e}_x$, $\mathbf{e}_y$, $\bar{\mathbf{p}}_1$), $\phi_2$ is the azimuthal angle of the $f_2$ momentum in the $Z_2^*$ rest frame formed by the vectors ($\mathbf{e}_x$, $- \mathbf{e}_y$, $- \bar{\mathbf{p}}_1$),
\begin{align}
\alpha_1 \equiv \boldsymbol \alpha \cdot \mathbf{e}_x, ~~ \alpha_2 \equiv \boldsymbol \alpha \cdot \mathbf{e}_y, ~~ \alpha_3 \equiv \boldsymbol \alpha \cdot \bar{\mathbf{p}}_1,
\end{align}
$\bar{\beta}_i \equiv \frac{\beta_i}{|\boldsymbol \beta|} ~(i = 1, 2, 3)$, $\boldsymbol \beta$ is the momentum of the fermion $f_2$ in the center-of-momentum frame of the particles $f_2$ and $\bar{f}_1$,
\begin{align}
\boldsymbol \beta & = \boldsymbol \alpha |_{\substack{a_1 \leftrightarrow a_2, \theta_1 \leftrightarrow \theta_2, \phi_1 \leftrightarrow \phi_2 \\ \mathbf{e}_y \to - \mathbf{e}_y, \bar{\mathbf{p}}_1 \to - \bar{\mathbf{p}}_1}} = \boldsymbol \alpha |_{k \rightarrow - k} & \notag \\
& = \mathbf{e}_x \frac{\sqrt{a_1} (2 E_2 + \sqrt{\tilde{a}_2}) \sin \theta_1 \cos \phi_1 + \sqrt{a_2} (2 E_1^{\prime} + \sqrt{\tilde{a}_2}) \sin \theta_2 \cos \phi_2}{4 (E_2 + E_1^{\prime} + \sqrt{\tilde{a}_2})} & \notag \\
& + \mathbf{e}_y \frac{\sqrt{a_1} (2 E_2 + \sqrt{\tilde{a}_2}) \sin \theta_1 \sin \phi_1 - \sqrt{a_2} (2 E_1^{\prime} + \sqrt{\tilde{a}_2}) \sin \theta_2 \sin \phi_2}{4 (E_2 + E_1^{\prime} + \sqrt{\tilde{a}_2})} + \frac{\bar{\mathbf{p}}_1}{8 (E_2 + E_1^{\prime} + \sqrt{\tilde{a}_2})} & \notag \\
& \times \Bigl ((m_X^2 - a_1 - a_2) (\cos \theta_1 - \cos \theta_2) - k (1 - \cos \theta_1 \cos \theta_2) & \notag \\
& + \frac{\sqrt{\tilde{a}_2}}{m_X} (- 2 k + (m_X^2 + a_1 - a_2) \cos \theta_1 - (m_X^2 + a_2 - a_1) \cos \theta_2) \Bigr), & \\
|\boldsymbol \beta| & = \frac{\sqrt{\tilde{a}_2}}{2}, &
\end{align}
\begin{align}
\beta_1 \equiv \boldsymbol \beta \cdot \mathbf{e}_x, ~~ \beta_2 \equiv \boldsymbol \beta \cdot (-\mathbf{e}_y), ~~ \beta_3 \equiv \boldsymbol \beta \cdot (-\bar{\mathbf{p}}_1),
\end{align}
\begin{align}
\tilde{A}_{\pm} \equiv \frac{A_{X \rightarrow Z_1^* Z_2^*} (\tilde{p}_1, \tilde{p}_2, \pm 1, \pm 1)} {g_Z}, \qquad \tilde{A}_0 \equiv \frac{A_{X \rightarrow Z_1^* Z_2^*} (\tilde{p}_1, \tilde{p}_2, 0, 0)} {g_Z},
\end{align}
\begin{align}
\tilde{A}_{\parallel} \equiv \frac{\tilde{A}_+ + \tilde{A}_-}{\sqrt{2}} = \sqrt{2} a_Z (\tilde{a}_1, \tilde{a}_2), \qquad
\tilde{A}_{\perp} \equiv \frac{\tilde{A}_+ - \tilde{A}_-}{\sqrt{2}} = \frac{2 \sqrt{2}}{m_X} |\mathbf{k}_1 + \mathbf{k}_2^{\prime}| c_Z (\tilde{a}_1, \tilde{a}_2),
\end{align}
\begin{align}
\eta_- & \equiv (\bar{\alpha}_1 - i \bar{\alpha}_2 )(\bar{\beta}_1 - i \bar{\beta}_2) & \notag \\
& = \frac{1}{4 \sqrt{\tilde{a}_1 \tilde{a}_2} (E_1 + E_2^{\prime} + \sqrt{\tilde{a}_1}) (E_2 + E_1^{\prime} + \sqrt{\tilde{a}_2})} \Biggl (a_1 (2 E_2 + \sqrt{\tilde{a}_2}) (2 E_2^{\prime} + \sqrt{\tilde{a}_1}) \sin^2 \theta_1 & \notag \\
& + a_2 (2 E_1^{\prime} + \sqrt{\tilde{a}_2}) (2 E_1 + \sqrt{\tilde{a}_1}) \sin^2 \theta_2 + \sqrt{a_1 a_2} \sin \theta_1 \sin \theta_2 \Bigl( (2 E_2^{\prime} + \sqrt{\tilde{a}_1}) (2 E_1^{\prime} + \sqrt{\tilde{a}_2}) e^{- i \phi} & \notag \\
& + (2 E_1 + \sqrt{\tilde{a}_1}) (2 E_2 + \sqrt{\tilde{a}_2}) e^{i \phi} \Bigr) \Biggr), &
\end{align}
\begin{align}
r_{\alpha \beta} \equiv \sqrt{\frac{(1 + \bar{\alpha}_3) (1 + \bar{\beta}_3)}{(1 - \bar{\alpha}_3) (1 - \bar{\beta}_3)}}.
\end{align}
Note that the dependence of expression (\ref{The fully differential width}) on $\phi_1$ and $\phi_2$ reduces to a dependence on $\phi_1 + \phi_2$ and in Eq.~(\ref{The fully differential width}) the latter sum has to be substituted by $\phi$.

\section{$\frac{d \Gamma} {d a}$ of a decay $X \to Z_1^* Z_2^* \to f_1 \bar{f}_1 f_2 \bar{f}_2$ with $f_1 \neq f_2$}
\label{Appendix: A formula for d Gamma over d a of the decay into non-identical fermions}

It follows from Eq.~(\ref{an explicit formula for d Gamma over d a for any decay X-> Z_1^* Z_2^* -> f_1 antif_1 f_2 antif_2}) that for any decay (\ref{X-> Z_1^* Z_2^* -> f_1 antif_1 f_2 antif_2, f_1 neq f_2})
\begin{align}
\label{a formula for d Gamma over d a for the decay into non-identical fermions}
\frac{d \Gamma} {d a} = \left. \frac{d \Gamma} {d a_2} \right |_{a_2 = a} = \left. \left (\int \limits_{0}^{(m_X - \sqrt{a_2})^2} da_1 \frac {d^2 \Gamma} {da_1 da_2} \right) \right |_{a_2 = a},
\end{align}
where the differential width $\frac {d^2 \Gamma} {da_1 da_2}$ is determined by Eq.~(8) from Ref.~\citen{Zagoskin:2016}.

If the functions $|a_Z (a_1, a_2)|$, $|b_Z (a_1, a_2)|$, $|c_Z (a_1, a_2)|$, and ${\rm Re} (a_Z^* (a_1, a_2) \, b_Z (a_1, a_2))$ are independent of $a_1$ and $a_2$, integration of $\frac {d^2 \Gamma} {da_1 da_2}$ in Eq.~(\ref{a formula for d Gamma over d a for the decay into non-identical fermions}) yields
\begin{align}
\label{an explicit formula for d Gamma over d a for the decay into non-identical fermions}
\frac{d \Gamma} {d a} =& \frac {\sqrt{2} G_F^3 m_Z^8 m_X } {2^{11} 3^3 \pi^5} (a_{f_1}^2 + v_{f_1}^2) (a_{f_2}^2 + v_{f_2}^2) \frac{1}{D (a)} & \notag \\
& \times \Biggl [ (1 - \alpha) \Bigl \{24 (-23 \alpha + 4 \eta + 1) |a_Z|^2 - 16 (2 \alpha^2 + (9 \eta + 17) \alpha + 3 \beta^2 - 9 \eta^2 - 3 \eta & \notag \\
& - 1) {\rm Re} (a_Z^* b_Z) + (3 \alpha^3 + (8 \eta - 45) \alpha^2 + (18 \eta^2 - 208 \eta - 45) \alpha - 6 (8 \eta + 1) \beta^2 - 6 \alpha \beta^2 + 48 \eta^3 & \notag \\
& + 18 \eta^2 + 8 \eta + 3) |b_Z|^2 + 64 \alpha (\alpha^2 + 2 (3 \eta + 17) \alpha + 6 \beta^2 - 18 \eta^2 + 6 \eta + 1) |c_Z|^2 \Bigr \} & \notag \\
& + 6 \ln \left (\frac{1}{\alpha} \right) \Bigl \{4 (12 \alpha^2 + 6 (1 - 4 \eta) \alpha - \beta^2 + 3 \eta^2) |a_Z|^2 + 8 (6 \alpha^2 - 3 \eta (\eta + 2) \alpha & \notag \\
& + \alpha \beta^2 - 2 \eta \beta^2 + 2 \eta^3) {\rm Re} (a_Z^* b_Z) + (30 \alpha^2 + \beta^4 + 10 \alpha \beta^2 - 30 \eta^2 \alpha - 10 \eta^2 \beta^2 + 5 \eta^4) |b_Z|^2 + 32 \alpha & \notag \\
& \times (-6 \alpha^2 + 3 (\eta^2 + 4 \eta - 2) \alpha + (4 \eta - 1) \beta^2 - \alpha \beta^2 + \eta^2 (3 - 4 \eta)) |c_Z|^2 \Bigr \} + s \cdot 3 \sqrt{2} & \notag \\
& \times \Bigl \{\left (\frac{1}{\beta} P_1 r_{+ \eta} - 4 P_2 r_{- \eta} \right) \ln \Bigl (\frac{1}{4 \alpha (\frac{m_Z^4}{m_X^4} + \beta^2)} & \notag \\
& \times (\alpha^2 \beta^2 + (\eta - 4)^2 \alpha^2 + 2 \alpha \beta^2 + 2 \eta (\eta - 4) \alpha + \beta^2 + \eta^2 - s \sqrt{2} (1 - \alpha) & \notag \\
& \times (\beta (\alpha + 1) r_{+ \eta} + ((\eta - 4) \alpha + \eta) r_{- \eta}) + (1 - \alpha)^2 \sqrt{(4 \alpha + \beta^2 - \eta^2)^2 + 4 \eta^2 \beta^2}) \Bigr) & \notag \\
& + 2 \left (4 P_2 r_{+ \eta} + \frac{1}{\beta} P_1 r_{- \eta} \right) (\pi - \arg ((\eta - 4) \alpha^2 + (- \eta^2 + 6 \eta - 4) \alpha - \alpha \beta^2 - \beta^2 & \notag \\
& + \eta (1 - \eta) + s \frac{1 - \alpha}{\sqrt{2}} (\beta r_{+ \eta} - \frac{m_Z^2}{m_X^2} r_{- \eta}) + i (1 - \alpha) (-\beta (1 - \alpha) & \notag \\
& + s \frac{\frac{m_Z^2}{m_X^2} r_{+ \eta} + \beta r_{- \eta}}{\sqrt{2}}))) \Bigr \} \Biggr], &
\end{align}
where $\alpha (a) \equiv \frac{a}{m_X^2}$, $\beta \equiv \frac{m_Z \Gamma_Z}{m_X^2}$, $\eta (a) \equiv 1 + \frac{a - m_Z^2}{m_X^2}$, in place of $s$ one may take either $1$ or $-1$ (this choice does not influence the dependence of $\frac{d \Gamma} {d a}$ on $a$),
\begin{align}
P_1 \equiv & 4 (12 \alpha^2 + 4 (2 - 3 \eta) \alpha - \beta^2 + \eta^2) |a_Z|^2 + 4 (8 \alpha^2 - 2 \eta (\eta + 2) \alpha - 3 \eta \beta^2 + 2 \alpha \beta^2 + \eta^3) {\rm Re} (a_Z^* b_Z) & \notag \\
& + ((4 \alpha + \beta^2)^2 + \eta^2 (\eta^2 - 8 \alpha - 6 \beta^2)) |b_Z|^2 - 32 \alpha (4 \alpha^2 + \alpha \beta^2 + (4 - 4 \eta - \eta^2) \alpha & \notag \\
& + (1 - 3 \eta) \beta^2 + \eta^2 (\eta - 1)) |c_Z|^2, & \notag \\
P_2 \equiv & 2 (6 \alpha - \eta) |a_Z|^2 + (4 (\eta + 1) \alpha + \beta^2 - 3 \eta^2) {\rm Re} (a_Z^* b_Z) + \eta (4 \alpha + \beta^2 - \eta^2) |b_Z|^2 & \notag \\
& - 8 \alpha (2 (\eta + 2) \alpha + \beta^2 + \eta (2 - 3 \eta)) |c_Z|^2, & \notag \\
r_{\pm \eta} \equiv & \sqrt{\sqrt{(4 \alpha + \beta^2 - \eta^2)^2 + 4 \eta \beta^2} \pm (4 \alpha + \beta^2 - \eta^2)}. &
\end{align}

We define the argument $\arg z$ of a complex number $z$ as follows:
\begin{align}
\label{a definition of arg(z)}
& \arg z = \arctan \frac{{\rm  Im}\,z}{{\rm  Re} \,z} + \pi n ({\rm  Re} \,z,  \, {\rm  Im}\,z) ~~~ \forall z \in C | {\rm  Re} \,z \neq 0, & \notag \\
& \arg z = \pi \left(\frac{1}{2} + \Theta (- {\rm  Im}\,z) \right) ~~ \forall z \in C | ({\rm   Re} \,z =0 ~~ {\rm and} ~~ {\rm   Im}\,z \neq 0), &
\end{align}
where $n (x, y) \equiv \Theta (-x) + 2 \, \Theta (x) \Theta (-y) ~~ \forall x \neq 0$,
\begin{align}
\Theta (x) \equiv 0 ~~ \forall x \in (-\infty, 0], ~~\Theta (x) \equiv 1 ~~ \forall x \in (0, +\infty).
\end{align}
According to definition (\ref{a definition of arg(z)}), $\arg z \in [0, 2 \pi)$.

\section{The definitions and explicit formulas for $\frac{d \Gamma}{d a}$ and $\frac{d \Gamma}{d \theta}$}
\label{Appendix: The definitions and explicit formulas for d Gamma over d a and d Gamma over d theta}

In this Appendix we propose some general definitions of the differential widths $\frac{d \Gamma}{d a}$ and $\frac{d \Gamma}{d \theta}$ for any decay (\ref{X-> Z_1^* Z_2^* -> f_1 antif_1 f_2 antif_2}), and show that the differential widths defined this way coincide with those defined in the standard fashion for decays (\ref{X-> Z_1^* Z_2^* -> f_1 antif_1 f_2 antif_2, f_1 neq f_2}) and (\ref{X-> Z_1^* Z_2^* -> f antif f antif}) separately. Therefore, the distributions presented in Fig.~\ref{Plots of the distributions of a, theta, and phi}a are general distributions defined for any decay (\ref{X-> Z_1^* Z_2^* -> f_1 antif_1 f_2 antif_2}) which are calculated for the decay into non-identical leptons and the distributions in Fig.~\ref{Plots of the distributions of a, theta, and phi}b are the same general distributions calculated for the decay into identical leptons. Thus, comparison of Fig.~\ref{Plots of the distributions of a, theta, and phi}a and Fig.~\ref{Plots of the distributions of a, theta, and phi}b is sensible thanks to the existence of the general definitions of $\frac{d \Gamma}{d a}$ and $\frac{d \Gamma}{d \theta}$.

\subsection{The differential width $\frac{d \Gamma}{d a}$}

We define the function $\frac{d \Gamma}{d a}$ as
\begin{align}
\label{a definition of 1 over Gamma d Gamma over d a}
\frac{1}{\Gamma} \frac{d \Gamma}{d a} \equiv \frac{1}{2} \frac{d P_a}{d a},
\end{align}
where $d P_a$ is the probability that in decay (\ref{X-> Z_1^* Z_2^* -> f_1 antif_1 f_2 antif_2}) there is a $Z$ boson whose squared invariant mass lies in an interval $[a, a + da]$. To derive an explicit formula for the distribution $\frac{1}{\Gamma}\frac{d \Gamma}{d a}$, we should recall that for decay (\ref{X-> Z_1^* Z_2^* -> f_1 antif_1 f_2 antif_2, f_1 neq f_2})
\begin{align}
\label{a relation between the fully differential width and the fully differential number of events for the decay to non-identical fermions}
\lim_{N \to \infty} \frac{1}{N} \frac{d^5 N_{f_1 \neq f_2}}{d^5 p} = \frac{1}{\Gamma} \frac{d^5 \Gamma} {d^5 p},
\end{align}
where $d^5 p \equiv d a_1 d a_2 d \theta_1 d \theta_2 d \phi$, $d^5 N_{f_1 \neq f_2}$ is the number of the decays (\ref{X-> Z_1^* Z_2^* -> f_1 antif_1 f_2 antif_2, f_1 neq f_2}) in which the squared invariant mass of $Z_1^*$ ($Z_2^*$) is in an interval $[a_1, a_1 + d a_1]$ ($[a_2, a_2 + d a_2]$), the polar angle of $f_1$ ($f_2$) lies in $[\theta_1, \theta_1 + d \theta_1]$ ($[\theta_2, \theta_2 + d \theta_2]$), and the azimuthal angle between the planes of the decays $Z_1^* \rightarrow f_1 \bar{f}_1$ and $Z_2^* \rightarrow f_2 \bar{f}_2$ is in an interval $[\phi, \phi + d \phi]$, among $N$ decays (\ref{X-> Z_1^* Z_2^* -> f_1 antif_1 f_2 antif_2, f_1 neq f_2}).

Eq.~(\ref{a relation between the fully differential width and the fully differential number of events for the decay to non-identical fermions}) is consistent with the fact that for any decay (\ref{X-> Z_1^* Z_2^* -> f_1 antif_1 f_2 antif_2})
\begin{align}
\label{a formula for Gamma of any decay X-> Z_1^* Z_2^* -> f_1 antif_1 f_2 antif_2}
\int \limits_{0}^{m_X^2} d a_1 \int \limits_{0}^{(m_X - \sqrt{a_1})^2} d a_2 \int \limits_{0}^{\pi} d \theta_1 \int \limits_{0}^{\pi} d \theta_2 \int \limits_{0}^{2 \pi} d \phi \frac{d^5 \Gamma} {d^5 p} = \Gamma,
\end{align}
because
\begin{align}
\int \limits_{0}^{m_X^2} d a_1 \int \limits_{0}^{(m_X - \sqrt{a_1})^2} d a_2 \int \limits_{0}^{\pi} d \theta_1 \int \limits_{0}^{\pi} d \theta_2 \int \limits_{0}^{2 \pi} d \phi \frac{d^5 N_{f_1 \neq f_2}} {d^5 p} = N.
\end{align}

Using Eqs.~(\ref{a definition of 1 over Gamma d Gamma over d a}) and (\ref{a relation between the fully differential width and the fully differential number of events for the decay to non-identical fermions}), we obtain that for decay (\ref{X-> Z_1^* Z_2^* -> f_1 antif_1 f_2 antif_2, f_1 neq f_2})
\begin{align}
\label{an explicit formula for 1 over Gamma d Gamma over d a for the decay into non-identical fermions}
\frac{1}{\Gamma}\frac{d \Gamma}{d a} & = \frac{1}{2} \lim_{N \to \infty} \frac{1}{N} \left ( \left. \frac{d N_{f_1 \neq f_2}} {d a_1} \right |_{a_1 = a} + \left. \frac{d N_{f_1 \neq f_2}} {d a_2} \right |_{a_2 = a} \right) = \frac{1}{2} \left ( \left. \frac{1}{\Gamma}\frac{d \Gamma}{d a_1} \right |_{a_1 = a} + \left. \frac{1}{\Gamma}\frac{d \Gamma}{d a_2} \right |_{a_2 = a} \right) & \notag \\
& = \left. \frac{1}{\Gamma}\frac{d \Gamma}{d a_1} \right |_{a_1 = a} = \left. \frac{1}{\Gamma}\frac{d \Gamma}{d a_2} \right |_{a_2 = a}, &
\end{align}
since if we neglect $m_{f_1}$ and  $m_{f_2}$, then $\left. \frac{d \Gamma}{d a_1} \right |_{a_1 = a} = \left. \frac{d \Gamma}{d a_2} \right |_{a_2 = a}$ (see Eq.~(8) in Ref.~\citen{Zagoskin:2016}).

For any decay (\ref{X-> Z_1^* Z_2^* -> f antif f antif})
\begin{align}
\label{a relation between the fully differential width and the fully differential number of events for the decay to identical fermions}
\lim_{N \to \infty} \frac{1}{N} \frac{d^5 N_{f_1 = f_2}}{d^5 p} = \frac{1}{\Gamma} \cdot 2 \frac{d^5 \Gamma} {d^5 p},
\end{align}
where $d^5 N_{f_1 = f_2}$ is the number of the decays (\ref{X-> Z_1^* Z_2^* -> f antif f antif}) in which there is a $Z$ boson $Z_1^*$ with a squared invariant mass lying in an interval $[a_1, a_1 + d a_1]$ and a $Z$ boson $Z_2^*$ whose squared invariant mass is in $[a_2, a_2 + d a_2]$, the polar angle of $f_1$ ($f_2$) lies in an interval $[\theta_1, \theta_1 + d \theta_1]$ ($[\theta_2, \theta_2 + d \theta_2]$), and the azimuthal angle between the planes of the decays $Z_1^* \rightarrow f_1 \bar{f}_1$ and $Z_2^* \rightarrow f_2 \bar{f}_2$ is in $[\phi, \phi + d \phi]$, among $N$ decays (\ref{X-> Z_1^* Z_2^* -> f antif f antif}). Note that while for decay (\ref{X-> Z_1^* Z_2^* -> f_1 antif_1 f_2 antif_2, f_1 neq f_2}) $Z_1^*$ ($Z_2^*$) is defined as the $Z$ boson decaying into $f_1 \bar{f}_1$ ($f_2 \bar{f}_2$), for decay (\ref{X-> Z_1^* Z_2^* -> f antif f antif}) the choice of $Z_1^*$ and $Z_2^*$ is arbitrary, which leads to the difference between the definitions of $d^5 N_{f_1 \neq f_2}$ and $d^5 N_{f_1 = f_2}$.

Eq.~(\ref{a relation between the fully differential width and the fully differential number of events for the decay to identical fermions}) accords with Eq.~(\ref{a formula for Gamma of any decay X-> Z_1^* Z_2^* -> f_1 antif_1 f_2 antif_2}) due to the fact that
\begin{align}
\label{an integral of d^5 N_f_1 = f_2 over d^5 p}
\int \limits_{0}^{m_X^2} d a_1 \int \limits_{0}^{(m_X - \sqrt{a_1})^2} d a_2 \int \limits_{0}^{\pi} d \theta_1 \int \limits_{0}^{\pi} d \theta_2 \int \limits_{0}^{2 \pi} d \phi \frac{d^5 N_{f_1 = f_2}} {d^5 p} = 2 N.
\end{align}
The ``2'' in the right-hand side of Eq.~(\ref{an integral of d^5 N_f_1 = f_2 over d^5 p}) emerges because of the double counting  during the integration of $\frac{d^5 N_{f_1 = f_2}} {d^5 p}$ on $a_1$ and $a_2$.

It follows from Eqs.~(\ref{a definition of 1 over Gamma d Gamma over d a}) and (\ref{a relation between the fully differential width and the fully differential number of events for the decay to identical fermions}) that for decay (\ref{X-> Z_1^* Z_2^* -> f antif f antif})
\begin{align}
\label{an explicit formula for 1 over Gamma d Gamma over d a for the decay into identical fermions}
\frac{1}{\Gamma}\frac{d \Gamma}{d a} = \frac{1}{2} \lim_{N \to \infty} \frac{1}{N} \left. \frac{d N_{f_1 = f_2}} {d a_1} \right |_{a_1 = a} = \frac{1}{2} \lim_{N \to \infty} \frac{1}{N} \left. \frac{d N_{f_1 = f_2}} {d a_2} \right |_{a_2 = a} = \left. \frac{1}{\Gamma}\frac{d \Gamma}{d a_1} \right |_{a_1 = a} = \left. \frac{1}{\Gamma}\frac{d \Gamma}{d a_2} \right |_{a_2 = a}.
\end{align}

Combining Eqs.~(\ref{an explicit formula for 1 over Gamma d Gamma over d a for the decay into non-identical fermions}) and (\ref{an explicit formula for 1 over Gamma d Gamma over d a for the decay into identical fermions}), we infer that in the approximation $m_{f_1} = m_{f_2} = 0$ for any decay (\ref{X-> Z_1^* Z_2^* -> f_1 antif_1 f_2 antif_2})
\begin{align}
\label{an explicit formula for d Gamma over d a for any decay X-> Z_1^* Z_2^* -> f_1 antif_1 f_2 antif_2}
\frac{d \Gamma}{d a} = \left. \frac{d \Gamma}{d a_1} \right |_{a_1 = a} = \left. \frac{d \Gamma}{d a_2} \right |_{a_2 = a}.
\end{align}

\subsection{The differential width $\frac{d \Gamma}{d \theta}$}

Analogously, we define the differential width $\frac{d \Gamma}{d \theta}$ as
\begin{align}
\label{a definition of 1 over Gamma d Gamma over d theta}
\frac{1}{\Gamma}\frac{d \Gamma}{d \theta} \equiv \frac{1}{2} \frac{d P_{\theta}}{d \theta},
\end{align}
where $d P_{\theta}$ is the probability that in decay (\ref{X-> Z_1^* Z_2^* -> f_1 antif_1 f_2 antif_2}) there is a fermion whose polar angle lies in an interval $[\theta, \theta + d \theta]$.

Eqs.~(\ref{a definition of 1 over Gamma d Gamma over d theta}) and (\ref{a relation between the fully differential width and the fully differential number of events for the decay to non-identical fermions}) yield that for any decay (\ref{X-> Z_1^* Z_2^* -> f_1 antif_1 f_2 antif_2, f_1 neq f_2})
\begin{align}
\frac{1}{\Gamma} \frac{d \Gamma}{d \theta} = \frac{1}{2} \lim_{N \to \infty} \frac{1}{N} \left ( \left. \frac{d N_{f_1 \neq f_2}} {d \theta_1} \right |_{\theta_1 = \theta} + \left. \frac{d N_{f_1 \neq f_2}} {d \theta_2} \right |_{\theta_2 = \theta} \right) = \frac{1}{2} \left ( \left. \frac{1}{\Gamma} \frac{d \Gamma}{d \theta_1} \right |_{\theta_1 = \theta} + \left. \frac{1}{\Gamma} \frac{d \Gamma}{d \theta_2} \right |_{\theta_2 = \theta} \right).
\end{align}

According to Eq.~(\ref{the fully differential width for the decays with f_1 different from f_2}), the differential width $\frac{d^2 \Gamma}{d \theta_1 d \theta_2}$ of decay (\ref{X-> Z_1^* Z_2^* -> f_1 antif_1 f_2 antif_2, f_1 neq f_2}) is invariant under the substitution $\theta_1 \to \theta_2$ and $\theta_2 \to \theta_1$ if $A_{f_1} = A_{f_2}$ (see Eq.~(\ref{a definition of A_f}) for the definition of the quantity $A_f$). That is why for decay (\ref{X-> Z_1^* Z_2^* -> f_1 antif_1 f_2 antif_2, f_1 neq f_2}) in the case $A_{f_1} = A_{f_2}$
\begin{align}
\left. \frac{d \Gamma}{d \theta_1} \right |_{\theta_1 = \theta} = \left. \frac{d \Gamma}{d \theta_2} \right |_{\theta_2 = \theta},
\end{align}
and therefore
\begin{align}
\label{an explicit formula for 1 over Gamma d Gamma over d theta for the decay into non-identical fermions}
\frac{1}{\Gamma} \frac{d \Gamma}{d \theta} = \left. \frac{1}{\Gamma} \frac{d \Gamma}{d \theta_1} \right |_{\theta_1 = \theta} = \left. \frac{1}{\Gamma} \frac{d \Gamma}{d \theta_2} \right |_{\theta_2 = \theta}.
\end{align}

We find from Eqs.~(\ref{a definition of 1 over Gamma d Gamma over d theta}) and (\ref{a relation between the fully differential width and the fully differential number of events for the decay to identical fermions}) that for decay (\ref{X-> Z_1^* Z_2^* -> f antif f antif})
\begin{align}
\label{an explicit formula for 1 over Gamma d Gamma over d theta for the decay into identical fermions}
\frac{1}{\Gamma}\frac{d \Gamma}{d \theta} = \frac{1}{2} \lim_{N \to \infty} \frac{1}{N} \left. \frac{d N_{f_1 = f_2}} {d \theta_1} \right |_{\theta_1 = \theta} = \frac{1}{2} \lim_{N \to \infty} \frac{1}{N} \left. \frac{d N_{f_1 = f_2}} {d \theta_2} \right |_{\theta_2 = \theta} = \left. \frac{1}{\Gamma} \frac{d \Gamma}{d \theta_1} \right |_{\theta_1 = \theta} = \left. \frac{1}{\Gamma} \frac{d \Gamma}{d \theta_2} \right |_{\theta_2 = \theta}.
\end{align}

Combination of Eqs.~(\ref{an explicit formula for 1 over Gamma d Gamma over d theta for the decay into non-identical fermions}) and (\ref{an explicit formula for 1 over Gamma d Gamma over d theta for the decay into identical fermions}) yields that for any decay (\ref{X-> Z_1^* Z_2^* -> f_1 antif_1 f_2 antif_2}) wherein $A_{f_1} = A_{f_2}$
\begin{align}
\label{an explicit formula for d Gamma over d theta for any decay X-> Z_1^* Z_2^* -> f_1 antif_1 f_2 antif_2}
\frac{d \Gamma}{d \theta} = \left. \frac{d \Gamma}{d \theta_1} \right |_{\theta_1 = \theta} = \left. \frac{d \Gamma}{d \theta_2} \right |_{\theta_2 = \theta}.
\end{align}

\section{The fully differential distribution of the decay $h \to Z_1^* Z_2^* \to 4 \ell$}
\label{Appendix: A formula for the fully differential distribution of the decay h to Z_1^* Z_2^* to 4 ell}

It follows from Eqs.~(\ref{a relation between the fully differential width and the fully differential number of events for the decay to identical fermions}), (\ref{a relation between the fully differential width and the fully differential number of events for the decay to non-identical fermions}), (\ref{the ATLAS limitations on m12, m34, eta_e, and eta_mu}), and (\ref{the CMS limitations on m12, m34, eta_e, and eta_mu}) that
\begin{align}
\label{the percentages of decays selected by ATLAS or CMS}
\lim_{N_k \to \infty} \frac{N_k^i}{N_k} = \frac{\Gamma_k^i}{\Gamma_k}, \qquad k = 4 e, 4 \mu, 2 e 2 \mu, \qquad i = {\rm ATLAS, CMS},
\end{align}
where $N_k^i$ is the number of the decays $h \to Z_1^* Z_2^* \to k$ selected by ATLAS or CMS, among $N_k$ decays $h \to Z_1^* Z_2^* \to k$,
\begin{align}
\Gamma_k \equiv \int \limits_{0}^{m_X^2} d a_1 \int \limits_{0}^{a_{2 \, max}} d a_2 \int \limits_{0}^{\pi} d \theta_1 \int \limits_{0}^{\pi} d \theta_2 \int \limits_{0}^{2 \pi} d \phi \, 2 \frac{d^5 \Gamma_k} {d^5 p},
\end{align}
$a_{2 \, max} \equiv {\rm Min} \left (a_1, (m_X - \sqrt{a_1})^2 \right)$, $\frac{d^5 \Gamma_k} {d^5 p}$ is the fully differential width of the decay $h \to Z_1^* Z_2^* \to k$ (see Eqs.~(\ref{The fully differential width}) and (\ref{the fully differential width for the decays with f_1 different from f_2})),
\begin{align}
& \Gamma_{4 e}^i \equiv \int \limits_{a_{1 \, min}^i}^{a_{1 \, max}^i} d a_1 \int \limits_{a_{2 \, min}^i}^{a_{2 \, max}} d a_2 \int \limits_{\theta_{e \, min}^i}^{\pi - \theta_{e \, min}^i} d \theta_1 \int \limits_{\theta_{e \, min}^i}^{\pi - \theta_{e \, min}^i} d \theta_2 \int \limits_{0}^{2 \pi} d \phi \, 2 \frac{d^5 \Gamma_{4 e}} {d^5 p}, & \notag \\
& \Gamma_{4 \mu}^i \equiv \int \limits_{a_{1 \, min}^i}^{a_{1 \, max}^i} d a_1 \int \limits_{a_{2 \, min}^i}^{a_{2 \, max}} d a_2 \int \limits_{\theta_{\mu \, min}^i}^{\pi - \theta_{\mu \, min}^i} d \theta_1 \int \limits_{\theta_{\mu \, min}^i}^{\pi - \theta_{\mu \, min}^i} d \theta_2 \int \limits_{0}^{2 \pi} d \phi \, 2 \frac{d^5 \Gamma_{4 \mu}} {d^5 p}, & \notag \\
& \Gamma_{2 e 2 \mu}^i \equiv \int \limits_{a_{1 \, min}^i}^{a_{1 \, max}^i} d a_1 \int \limits_{a_{2 \, min}^i}^{a_{2 \, max}} d a_2 \int \limits_{\theta_{e \, min}^i}^{\pi - \theta_{e \, min}^i} d \theta_1 \int \limits_{\theta_{\mu \, min}^i}^{\pi - \theta_{\mu \, min}^i} d \theta_2 \int \limits_{0}^{2 \pi} d \phi \, 2 \frac{d^5 \Gamma_{2 e 2 \mu}} {d^5 p}, &
\end{align}
\begin{align}
& a_{1 \, min}^{{\rm ATLAS}} = (50 \, {\rm GeV})^2, \qquad a_{1 \, max}^{{\rm ATLAS}} = (106 \, {\rm GeV})^2, \qquad a_{2 \, min}^{{\rm ATLAS}} = (12 \, {\rm GeV})^2, & \notag \\
& \theta_{e \, min}^{{\rm ATLAS}} \equiv 2 \arctan e^{-2.47}, \qquad  \theta_{\mu \, min}^{{\rm ATLAS}} \equiv 2 \arctan e^{-2.7}, &
\end{align}
\begin{align}
& a_{1 \, min}^{{\rm CMS}} = (40 \, {\rm GeV})^2, \qquad a_{1 \, max}^{{\rm CMS}} = (m_X - 12 \, {\rm GeV})^2, \qquad a_{2 \, min}^{{\rm CMS}} = (12 \, {\rm GeV})^2, & \notag \\
& \theta_{e \, min}^{{\rm CMS}} \equiv 2 \arctan e^{-2.5}, \qquad \theta_{\mu \, min}^{{\rm CMS}} \equiv 2 \arctan e^{-2.4}.
\end{align}

Moreover, the fully differential distribution of the decay $h \to Z_1^* Z_2^* \to 4 \ell$ is
\begin{align}
\label{some formulas for the fully differential distribution of the decay h to Z1 Z2 to 4 ell}
\lim_{N_{4 \ell} \to \infty} \frac{1}{N_{4 \ell}} \frac{d^5 N_{4 \ell}}{d^5 p} & = \lim_{N_{4 \ell} \to \infty} \frac{1}{N_{4 e} + N_{4 \mu} + N_{2 e 2 \mu}} \frac{d^5 N_{4 e} + d^5 N_{4 \mu} + d^5 N_{2 e 2 \mu}^{\prime}} {d^5 p} & \notag \\
& = \lim_{N_{4 \ell} \to \infty} \frac{1}{2 N_{4 e} + N_{2 e 2 \mu}} \frac{2 d^5 N_{4 e} + d^5 N_{2 e 2 \mu}^{\prime}}{d^5 p}, &
\end{align}
where
\begin{itemize}
\item $d^5 N_{4 \ell}$ is the number of the decays $h \to Z_1^* Z_2^* \to 4 \ell$ in which $m_{12}^2 \in [a_1, a_1 + d a_1]$, $m_{34}^2 \in [a_2, a_2 + d a_2]$, the polar angle of $f_1$ ($f_2$) lies in an interval $[\theta_1, \theta_1 + d \theta_1]$ ($[\theta_2, \theta_2 + d \theta_2]$), and the azimuthal angle between the planes of the decays $Z_1^* \rightarrow f_1 \bar{f}_1$ and $Z_2^* \rightarrow f_2 \bar{f}_2$ is in $[\phi, \phi + d \phi]$, among $N_{4 \ell}$ decays $h \to Z_1^* Z_2^* \to 4 \ell$;
\item $d^5 N_{2 e 2 \mu}^{\prime}$ is the number of the decays $h \to Z_1^* Z_2^* \to 2 e 2 \mu$ in which $m_{12}^2 \in [a_1, a_1 + d a_1]$, $m_{34}^2 \in [a_2, a_2 + d a_2]$, the polar angle of $f_1$ ($f_2$) lies in an interval $[\theta_1, \theta_1 + d \theta_1]$ ($[\theta_2, \theta_2 + d \theta_2]$), and the azimuthal angle between the planes of the decays $Z_1^* \rightarrow f_1 \bar{f}_1$ and $Z_2^* \rightarrow f_2 \bar{f}_2$ is in $[\phi, \phi + d \phi]$, among $N_{4 \ell}$ decays $h \to Z_1^* Z_2^* \to 4 \ell$.
\end{itemize}

Hereinafter, the symbol $Z_1^*$ ($Z_2^*$) denotes the $Z$ boson whose mass is $m_{12}$ ($m_{34}$) and $f_1$ ($f_2$) denotes the fermion whose parent $Z$ boson is $Z_1^*$ ($Z_2^*$).

It follows from Eqs.~(\ref{a relation between the fully differential width and the fully differential number of events for the decay to non-identical fermions}) and (\ref{the fully differential width for the decays with f_1 different from f_2}) that
\begin{align}
\label{d^5 N_2 e 2 mu^prime = 2 d^5 N_2 e 2 mu}
d^5 N_{2 e 2 \mu}^{\prime} = 2 \, d^5 N_{2 e 2 \mu}.
\end{align}

Using Eqs.~(\ref{some formulas for the fully differential distribution of the decay h to Z1 Z2 to 4 ell}) and (\ref{d^5 N_2 e 2 mu^prime = 2 d^5 N_2 e 2 mu}), we derive that
\begin{align}
\label{an explicit formula for the fully differential distribution of the decay h to Z1 Z2 to 4 ell}
\lim_{N_{4 \ell} \to \infty} \frac{1}{N_{4 \ell}} \frac{d^5 N_{4 \ell}}{d^5 p} & = \lim_{N_{4 \ell} \to \infty} \frac{2}{2 \Gamma_{4 e} + \Gamma_{2 e 2 \mu}} \left ( \frac{\Gamma_{4 e}}{N_{4 e}} \frac{d^5 N_{4 e}}{d^5 p} + \frac{\Gamma_{2 e 2 \mu}}{N_{2 e 2 \mu}} \frac{d^5 N_{2 e 2 \mu}}{d^5 p} \right) & \notag \\
& = \frac{2}{2 \Gamma_{4 e} + \Gamma_{2 e 2 \mu}} \left ( 2 \frac{d^5 \Gamma_{4 e}}{d^5 p} + \frac{d^5 \Gamma_{2 e 2 \mu}}{d^5 p} \right). &
\end{align}

Integration of Eq.~(\ref{an explicit formula for the fully differential distribution of the decay h to Z1 Z2 to 4 ell}) yields
\begin{align}
\label{the percentage of decays h to ZZ to ell selected by ATLAS or CMS}
\lim_{N_{4 \ell} \to \infty} \frac{N_{4 \ell}^i}{N_{4 \ell}} = \frac{2 \Gamma_{4 e}^i + \Gamma_{2 e 2 \mu}^i}{2 \Gamma_{4 e} + \Gamma_{2 e 2 \mu}},
\end{align}
where $N_{4 \ell}^i$ is the number of the decays $h \to Z_1^* Z_2^* \to 4 \ell$ selected by ATLAS or CMS, among $N_{4 \ell}$ decays $h \to Z_1^* Z_2^* \to 4 \ell$.

Besides, we obtain from Eq.~(\ref{an explicit formula for the fully differential distribution of the decay h to Z1 Z2 to 4 ell}) that
\begin{align}
\label{an explicit formula for the fully differential distribution of the decay h to Z1 Z2 to 4 ell with the ATLAS or CMS limitations}
\lim_{N_{4 \ell} \to \infty} \frac{1}{N_{4 \ell}^i} \frac{d^5 N_{4 \ell}}{d^5 p} & = \frac{2}{2 \Gamma_{4 e} + \Gamma_{2 e 2 \mu}} \left ( 2 \frac{d^5 \Gamma_{4 e}}{d^5 p} + \frac{d^5 \Gamma_{2 e 2 \mu}}{d^5 p} \right) \lim_{N_{4 \ell} \to \infty} \frac{N_{4 e} + N_{4 \mu} + N_{2 e 2 \mu}}{N_{4 e}^i + N_{4 \mu}^i + N_{2 e 2 \mu}^i} & \notag \\
& = \frac{2}{\Gamma_{4 e}^i + \Gamma_{4 \mu}^i + \Gamma_{2 e 2 \mu}^i} \left ( 2 \frac{d^5 \Gamma_{4 e}}{d^5 p} + \frac{d^5 \Gamma_{2 e 2 \mu}}{d^5 p} \right). &
\end{align}
%



\end{document}